\documentclass[floatfix, preprint2]{aastex6}

\usepackage{epsfig}
\usepackage{graphicx}
\usepackage{longtable}
\usepackage{natbib}
\usepackage{subfigure}
\usepackage{xcolor}
\usepackage{amsmath}
\usepackage{multirow}
\usepackage{float}

\newcommand{\Msolar}{M$_{\odot}$}

\newcommand{\Rsolar}{R$_{\odot}$}
\newcommand{\kms}{km s$^{-1}$}

\shorttitle{Post-interaction Stars on the M67 Main Sequence}
\shortauthors{Leiner et al.}

\begin{document}

\title{Blue Lurkers:  Hidden Blue Stragglers on the M67 Main Sequence Identified from their Kepler/K2 Rotation Periods}

\author{Emily Leiner\altaffilmark{1, 2, 7}, Robert
  D. Mathieu\altaffilmark{2}, Andrew Vanderburg\altaffilmark{3,8}, Natalie M. Gosnell\altaffilmark{4} and Jeffrey C. Smith\altaffilmark{5,6}}
\email{emily.leiner@northwestern.edu}

\altaffiltext{1}{Center for Interdisciplinary Exploration and Research in Astrophysics, Northwestern University, 2145 Sheridan Rd, Evanston, IL 60208, USA}
\altaffiltext{2}{Department of Astronomy, University of Wisconsin-Madison, 475 North Charter St, Madison, WI 53706, USA}
\altaffiltext{3}{Department of Astronomy, University of Texas at Austin, Austin, TX 78712, USA}
\altaffiltext{4}{Department of Physics, Colorado College, 14 E. Cache La Poudre St, Colorado Springs, CO  80903, USA}
\altaffiltext{5}{SETI Institute, Mountain View, CA 94043, USA} 
\altaffiltext{6}{NASA Ames Research Center, Moffett Field, CA, 94035, USA}
\altaffiltext{7}{NSF AAPF Fellow}
\altaffiltext{8}{NASA Sagan Fellow}

\begin{abstract}
 At an age of 4 Gyr, typical solar-type stars in M67 have rotation rates of 20-30 days. Using K2 Campaign 5 and 16 light curves and the spectral archive of the WIYN Open Cluster Study, we identify eleven three-dimensional kinematic members of M67 with anomalously fast rotation periods of 2-8 days, implying ages of less than 1 Gyr.  We hypothesize that these anomalously fast rotators have been spun up by mass transfer, mergers, or stellar collisions during dynamical encounters within the last Gyr, and thus represent lower-luminosity counterparts to the blue straggler stars. These 11 candidate post-interaction stellar systems have much in common with the blue stragglers including a high binary fraction (73\%), a number of long-period, low-eccentricity binary systems, and in at least one case a UV excess consistent with the presence of a hot white dwarf companion. The identification of these 11 systems provides the first picture of the low-luminosity end of the blue straggler distribution, providing new constraints for detailed binary evolution models and cluster population studies. This result also clearly demonstrates the need to properly account for the impact of binaries on stellar evolution, as significant numbers of post-interaction binaries likely exist on cluster main sequences and in the field. These stars are not always easy to identify, but make up $\sim10$\% or more of the spectroscopic binary population among the solar-type stars in M67.

\end{abstract}
 \section{Introduction}\label{section:intro}

In color-magnitude diagrams (CMDs) of star clusters, blue straggler stars (BSSs) are found brighter and bluer than the main sequence turnoff. BSSs are thought to form from mass transfer in binary systems \citep{McCrea1964, Gosnell2014}, stellar collisions during dynamical encounters \citep{Leonard1989, Sills2001}, or binary mergers (e.g. induced by Kozai cycles \citealt{Perets2009, Ivanova2008}). 

Blue stragglers are not the only mass-transfer, merger, or collision products that exist in clusters. Evolved counterparts to the blue stragglers (sometimes called `yellow giants' or 'yellow stragglers') are observed in between the blue straggler region and the red giant branch or detected as over-massive cluster giants via asteroseismology\citep{Landsman1997, Leiner2016, Handberg2017, Corsaro2012}. 

In principle, lower-mass blue stragglers could form via mass accretion onto initially lower-mass secondaries, through less efficient mass-transfer processes, or via mergers or collisions of two lower-mass main-sequence stars.  Such lower-mass blue stragglers would be hidden within cluster main sequences. Indeed, N-body and population synthesis studies predict that such mass transfer or merger products may be numerous \citep{Andronov2006, Geller2013}.

Actually detecting these low-mass blue stragglers on the main sequence is challenging. Very close main-sequence-white dwarf (MS-WD) binaries can be detected in time-series photometric surveys if they are eclipsing (e.g. \citealt{Parsons2015, Almenara2012, Breton2012}, or from X-ray and transient surveys in cases where there is active accretion and/or outbursts (i.e. novae and cataclysmic variables) (e.g. \citealt{Fornasini2014, Strope2010, Szkody2011}). Post-mass-transfer binaries in wider orbits (P$> 10$ days) with hot white dwarf companions  can also be identified in UV surveys (e.g. \citealt{Gosnell2014, Jeffries1996, vanRoestel2018, Parsons2016,  Li2014, Rebassa-Mansergas2010, Rebassa-Mansergas2017}), but older post-mass transfer systems with fainter, cooler white dwarf companions escape detections in these studies, as do merger and collision products. As a result, the full extent of the post-interaction main-sequence population of clusters is not known. A better census of the post-interaction population requires developing other techniques that may be used to identify these post-interaction stars that blend photometrically with cluster main-sequences. 

One method is using stellar abundance measurements. Stellar merger and mass-transfer products are predicted to have spectral signatures including barium, carbon, oxygen, or lithium abundance anomalies. This technique has identified many blue straggler counterparts in the field like carbon-enhanced metal poor stars with s-process enrichment (CEMP-s stars), barium stars, and lithium enhanced giants  (e.g. \citealt{Jorissen1998,Hansen2016, Aoki2008}). Detecting these post-interaction systems from abundance signatures requires high-resolution spectra, and known blue stragglers do not always have the observed abundance signatures expected from mass-transfer or collisional formation \citep{Shetrone2000, Milliman2016}. The observational biases and completeness of abundance detection methods are not well defined. 

Here we propose an alternative technique to identify recent mass-transfer and collision products, using rotation rates. Recent advances in our understanding of stellar angular momentum evolution have revealed a clearer picture of the rotational evolution of solar-like stars. Observations of young clusters show that on the pre-main-sequence these stars have a wide range of rotation periods. Early in their lives these stars spin down due to magnetic braking (e.g. via a magnetized wind or disc locking \citealt{Gallet2013, Matt2005}), with faster rotators spinning down more quickly due to their stronger magnetic field. After several hundred Myr, solar-type stars of the same age will converge to the same rotation rate regardless of their initial angular momentum (e.g. \citealt{Barnes2003, Gallet2013, Meibom2015, Meibom2009, Epstein2014}). Thereafter stellar rotation rates can be used as a proxy for stellar age, a technique known as gyrochronology. Recently, \citet{Leiner2018} have suggested the same gyrochronology age determinations may also be used in post-mass-transfer systems. 

Given the wealth of observational and theoretical evidence (Section~\ref{section:rotationalevolution}), we assume that anomalously rapid rotation rates  are observed among all stars that have recently undergone a merger, collision, or mass-transfer event. Therefore, rotation rates from spectroscopic $v$sin$i$ measurements or photometric spot modulation may be an effective way to select for recent stellar interaction products. As a test case, we look at stellar rotation rates among stars in the old (4 Gyr) open cluster M67, looking for any main-sequence cluster members with rotation periods much shorter than the 20-30 days measured for main sequence stars in the cluster\footnote{We note that tidally synchronized binaries will be rotating faster than this. These stars are easily excluded from our sample as we explain in Section~\ref{section:rotmeasurements}} \citep{Barnes2016, Gonzalez2016}. This study is the first to use rotation to identify the post-interaction population of a cluster, and offers the first glimpse of the low-luminosity end of the blue straggler distribution.

In Section 2 we discuss our premise that rapid rotation is a sign of mass-transfer, merger or collision formation. In Section 3 we discuss the K2 observations of M67 and our technique for light-curve extraction and analysis. In Section 4 we discuss each of our candidate post-interaction systems in detail.  In Section 5 we discuss the overall population characteristics of the anomalously rapid rotators in the cluster that we suggest formed from recent mass transfer or collision events. In Section 6 we discuss the significance of these detections and summarize our results.

 \section{The Rotational Evolution of Mass-transfer Products}\label{section:rotationalevolution}

In theory, mass transfer in a binary also transports significant angular momentum, resulting in substantial spin-up of the mass accreting star \citep{Packet1981, DeMink2013}. Similarly, stellar collisions are expected to yield rapidly rotating stellar products \citep{Sills2002, Sills2005}. These interactions, then, can be seen as resetting the gyro-age clock, giving old stars the rapid rotation rates indicative of youth.  

Observations confirm that many mass-transfer and collision products like the blue stragglers are rotating rapidly (e.g. \citealt{Carney2005, Nemec2017, Jeffries1996, Mucciarelli2014, Lovisi2010}). These blue stragglers are sometimes observed to have $v$sin$i$ measurements as large as 200 \kms. In previous work \citep{Leiner2018}, we provided the first observational study of spin-down in post-mass-transfer binaries. This work used a sample of 12 post-mass-transfer systems, all composed of an FGK main-sequence star with a hot white dwarf companion. The white dwarfs in these systems had measured temperatures, and thus their ages (i.e. time since mass transfer ceased) could be determined from white dwarf cooling. The FGK primaries also had measured rotation rates from either photometric spot modulation or spectroscopic vsinis. Comparing the white dwarf cooling ages to measured rotation periods, \citet{Leiner2018} concluded that young (several Myr) post-mass-transfer systems have rotation periods $< 1$ day, or $30-40\%$ of their critical rotation rate. Older systems have slower rotation rates, with the relationship following approximately the spin-down curves for normal solar-type stars from \citet{Gallet2015}. From this work, \citet{Leiner2018} concluded that rotation may be a useful indicator for age in post-mass-transfer systems in which white dwarf ages are not available. Further, they suggested that gyrochronology relationships developed for normal FGK stars (i.e. \citealt{Angus2015}) may also be applicable to FGK post-mass-transfer binaries. 

Another implication of this work is that rapid rotation rates may be indicative of a recent episode of mass transfer. In M67, an old (4 Gyr) open cluster, typical main-sequence rotation rates are 20-30 days \citep{Gonzalez2016, Barnes2016}, except for systems in short-period binaries in which rotation has been tidally synchronized with the orbital period. Much shorter rotation periods, then, may be a way to select systems that have been through a recent interaction. Here we test this idea in M67, searching for anomalously rapid rotators in the cluster that may be post-interaction binaries formed in mass-transfer, mergers, or collisions.

\section{Observations}\label{section:observations}
\subsection{WIYN Open Cluster Study}\label{section:WIYN}
M67 is a well-studied old (4 Gyr) open cluster. It has extensive archival photometry \citep{Montgomery1993, Fan1996,vandenBerg2004}, proper-motion memberships \citep{Sanders1977, Girard1989}, and radial-velocity membership information from more than 40 years of high-precision radial velocities obtained on the WIYN 3.5m telescope with the Hydra Multi-Object Spectrograph and with the Harvard-Smithsonian Center for Astrophysics Digital Speedometers. These radial-velocity data are stored in the archive of the WIYN Open Cluster Study (WOCS; \citep{Mathieu2000}). \citet{Geller2015} incorporate both proper motions and radial velocities to determine memberships and binary status for stars in the cluster field out to a radius of 30$\arcmin$. A subsequent paper will publish orbital solutions for the known binaries (Geller et al. 2019, in prep). This WOCS synthesis contains members down to a limiting magnitude of V= 16.5, a sample that includes blue stragglers, the subgiant and giant branches, and FGK main-sequence stars. In our analysis, we adopt the most up-to-date membership information and binary orbital parameters from these WOCS papers. 

\subsection{K2 Observations of M67}\label{section:K2}
M67 was observed in Campaign 5 (K2 Guest Observer Program 5031, Mathieu, PI) of the \textit{Kepler} space telescope's extended K2 mission for 76 days between 27 April and 10 July 2015, using a combination of individual apertures and a 25' by 25' superstamp of pixels covering the cluster center. M67 was reobserved in Campaign 16 (7 December 2017 -- 25 February 2018), providing an additional 81 days of time-series photometry for most of the cluster members observed in Campaign 5 in addition to light curves for some new sources not observed in the first campaign. Campaign 18 (12 May 2018-- 2 July 2018) also reobserved the Campaign 5 field, but was cut short due to low fuel on board the spacecraft and thus yielded 51 day light curves for all targets. Data products for the C18 superstamp covering most of M67 are not available at the time of this analysis, and so we do not include them here. Light curves for individual targets are available for C18, but only one of the stars analyzed in this paper falls into this category. For Campaign 5 and 16, light curves for both the individual targets and the targets in the superstamp were extracted and corrected for K2 systematic errors using the method of \citet{Vanderburg2014} and \citet{Vanderburg2016}. While these methods effectively remove systematics caused by the K2 6-hour pointing drift, they leave in long-term instrumental systematics which can impede searches for long-period signals like stellar rotation. To remove these systematics, we used the \textit{Kepler} team's Pre-search Data Conditioning-Maximum A Posteriori (PDC-MAP) software \citep{Stumpe2014, Smith2012} to identify and remove common-mode instrumental trends. Unlike the standard \textit{Kepler} export data products, we utilized \textit{single-scale} PDC-MAP, which performs best at removing long term trends while preserving long-period signals. This process is described in more detail in \citep{Esselstein2018}. 

\subsection{Rotation Measurements}\label{section:rotmeasurements}
\begin{figure}
\vspace{.5cm}
\includegraphics[angle=0, width= .9\linewidth ]{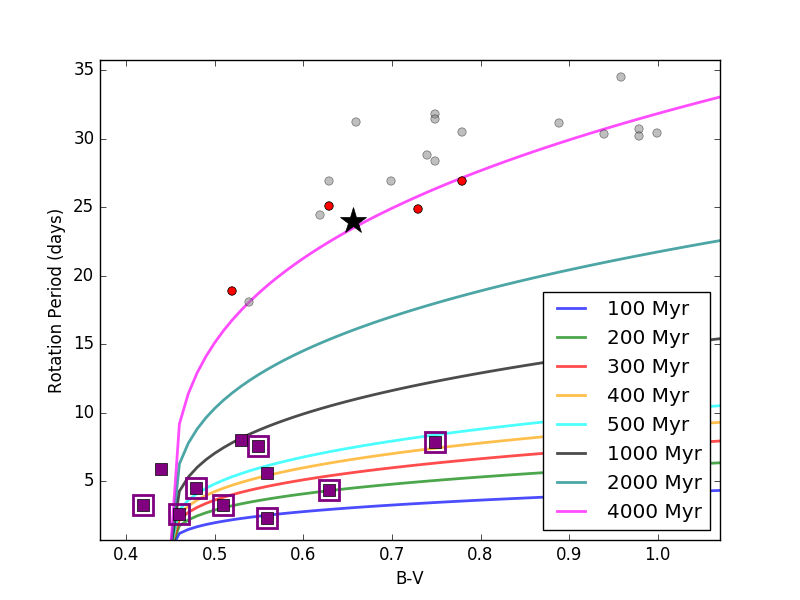}
\caption{A color-rotation plot comparing the 11 rapid rotators in our sample (purple points) to a sample of normal M67 main-sequence stars with rotation periods from \citet{Barnes2016} (gray points). All are de-reddened using E(B-V)= 0.041 \citep{Taylor2007}. We highlight in red the \citet{Barnes2016} rotation periods that also have a measurement from the CfA and/or Oxford pipeline that agrees within $15\%$. Binaries are boxed as in Figure \ref{CMD}. For comparison, we also show gyrochronology models from \citet{Angus2015} for ages ranging from 100 Myr to 4 Gyr and the rotation rate of the Sun (black star).}\label{fig:gyro}
\end{figure}

\begin{figure}
\vspace{.5cm}
d\includegraphics[angle=0, width= .9\linewidth ]{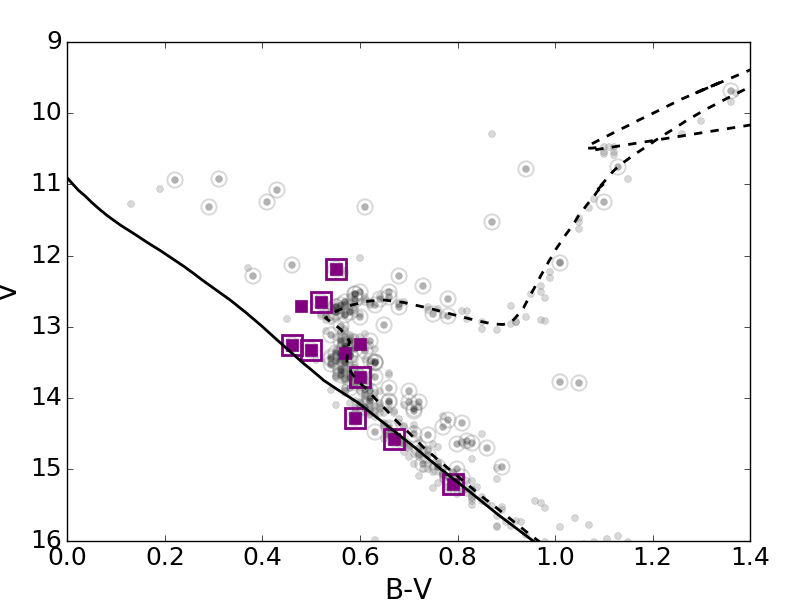}
\caption{A color-magnitude diagram showing 3D kinematic members of M67, with binary members boxed/circled \citep{Geller2015}. Purple points show stars with higher-than-normal rotation rates, which we suggest are products of recent stellar interactions. The black solid and dashed lines are the the ZAMS and a 4 Gyr isochrone, respectively. }\label{CMD}
\end{figure}

\begin{table*}
\centering
    \caption{Stellar and Orbital Properties of Rapid Rotators}
		\begin{tabular}{lcccccccc}
		\hline
		EPIC ID& WOCS ID & Orbital Period & Eccentricity & f(m) &  V & B-V & Rotation Period & Membership\tablenote{\scriptsize{Membership classification are explained in \citet{Geller2015}. BM= Binary member, BLM= Binary Likely Member, SM= Single Member}} \\
        & & (days) & & (\Msolar) & & & (days) & \\
		\hline
	211411716 & 4001 & 139.77 & 0.36 & 2.17e-2 &15.21& 0.79 & 7.9 & BM\\
	211410278 & 14020 & 358.9   & 0.23 & 2.38e-3 &14.58 & 0.67	& 4.4  & BM\\
	211421624 &12020 & 762   & 0.056 & 2.87e-2 &14.28& 0.59 & 7.6 & BM\\
+	211411928 & 3001  & 128.14   & 0.04 & 1.43e-2 & 13.26 & 0.46 & 3.3\tablenote{\scriptsize{These stars are included based on their $v$sin$i$ measurements ($v$sin$i > 10$ \kms), not periodic signals in their K2 light curves.}} & BM\\
	211428722 & 2068 & 8567 & 0.859  & 6.81e-2 & 12.19 & 0.55 & 20 \& 3 & BM \\
	211409959 & 9005 & 2769  & 0.15 & 3.68e-2 & 12.65 & 0.52 & 4.5 & BM\\
	211404255 & 6025 & 6265  & 0.38 & 2.20e-1 & 13.70 & 0.6 & 2.3 & BM\\
	211414427 &11006 & $>  3500$ & \nodata& \nodata &13.33 & 0.50 & 2.6\tablenotemark{b}& BLM\\
	211406971 & 1020 & \nodata & \nodata & \nodata &12.70 & 0.48 & 5.9 & SM\\
	211411690 & 2001& \nodata  & \nodata & \nodata & 13.24 & 0.60& 5.6 days& SM\\
    211427425 & 7035 & \nodata & \nodata & \nodata & 13.36 & 0.57& 8.0  & SM \\
	\hline
	\end{tabular}
	\label{table}
\end{table*}

We selected all three-dimensional (3D) kinematic members or likely members of the M67 main sequence observed in K2 Campaign 5 or 16. For each star, we created a Lomb-Scargle periodogram using the light--curve processing software \texttt{vartools} \citep{Hartman2008}. As a first cut to remove power spectra without periodic signals, we selected all stars from this sample with measured periods less than 15 days and power of at least 0.1. We cross-referenced these stars with the binary orbital information from WOCS (Geller et al. 2019, in prep) to exclude any short-period binaries with orbital periods less than 60 days. We do this because  tidal forces spin up the rotation rates of close binaries, explaining any observed rapid rotation. Among the $\sim2800$ field eclipsing binaries in the Kepler Eclipsing Binary Catalog, \citet{Lurie2017} find the fraction of tidally circularized binaries drops off at periods greater than 10 days, and the fraction of tidally synchronized binaries drops off dramatically at periods longer than 30 days. These cutoff periods are also compatible with tidal circularization studies  in open clusters \citep{Meibom2005, Meibom2009}. As a conservative cut, we double the \citet{Lurie2017} tidal synchronization limit, removing any binaries with $P_\text{orb} < 60$ days from our sample.

We compare the remaining sample to the rotational models of \citet{Angus2015}, selecting all stars with rotation rates faster than the 1 Gyr model (Figure~\ref{fig:gyro}). This age cut allows us to take into account the temperature of the star when determining if rotation rates are unusual, as bluer stars close to the cluster turnoff naturally have slightly faster rotation rates than redder stars further down the main sequence. These rotational models are undefined for stars hotter than $(B-V)_0$= 0.45. There are a few stars in M67 blueward of this limit, and so for these hotter stars we use a rotation cut of $P{_{rot} }> 8.0$ day, the approximate rotation rate of a 1 Gyr star near the cluster turnoff. For context, in Figure ~\ref{fig:gyro} we also show the rotation periods of a sample of normal main-sequence stars in M67 from \citet{Barnes2016} (gray points). These stars have rotation periods of $\sim25$ days. We note that \citet{Esselstein2018} raise doubts about the reliability of many of the published M67 rotation periods, including those of \citet{Barnes2016}. They find that for many sources, different pipelines yield different rotation measurements. To highlight a more reliable sample, we show in red the rotation periods from \citet{Barnes2016} that also have a rotation measurement in \citet{Esselstein2018} that agrees within $15\%$. We also show the rotation period of the Sun, which is close in age to M67 stars.

For our sample of fast rotating stars, we visually examine all the light curves and periodograms to remove any spurious or marginal results. We exclude some lower signal-to-noise systems with multiple peaks. We check the light curves of each target's neighboring stars in the EPIC catalog within 30 arcseconds of each target to determine if the observed periodic signals might originate with a nearby variable star. We also visually examine the CCD images from K2 to check for nearby stars, and cross reference with the 2MASS catalog to check for any stars within 30 arcseconds that may be missing from the EPIC database or too faint to identify in the images. In addition, we adjusted the size of the photometric aperture used to extract the light curve to check that the variability appears to be centered on these sources, and does not become stronger with a larger aperture. Using these techniques we remove several systems where the variability appears to originate with a neighboring star. These steps give us confidence that the remaining stars in our samples are true rotational variables.

We find 9 stars in our sample that show rotation rates much faster than those of normal main-sequence stars in M67. We show the CMD location of these stars in Figure~\ref{CMD}. We also show raw light curves, phase-folded light curves, and Lomb-Scargle periodograms from both C5 and C16 for these 9 stars in Figure~\ref{LC}. The measurement precision of these periods is generally good to a few percent for periods of several days, up to 10-20\% for the longer 20-day periods or multi-periodic sources in our sample. We do not quote these measurement errors because for most of our sources they are misleadingly small. The more significant sources of error will be astrophysical, such as spot migration and differential rotation. We expect these to cause typical rotation period variations on the order of 10\%, though in some stars (with more extreme differential rotation, for example) it may be higher \citep{Reinhold2015, Lurie2017, Balona2016}. 

As an additional independent check, we also compare these rotation periods to those produced using another light curve production pipeline. Esselstein et al. 2018 compare periods measured from the \citet{Vanderburg2014} light curves we use here and the Oxford pipeline light curves of \citet{Aigrain2015}. For all sources, our measured periods agree with the Esselstein et al. period measurements from the Vanderburg light curves and with the Esselstein et al. period measurements from the Oxford pipeline, though four sources (WOCS 2068, 1020, 7035, and 9005) do not meet their more conservative detection criteria. We classify these detections as less certain and discuss these cases in more detail in the next section.  

In addition, we measure $v$sin$i$ rates for all 3D kinematic members of the cluster from \citet{Geller2015} (see \citealt{Geller2008} for an explanation of our $v$sin$i$ measurement technique), again excluding short-period binaries from our sample. The WOCS spectra have a  $v$sin$i$ measurement limit of 10 \kms. Typical stars in M67 with rotation rates of 20-30 days would be rotating with surface velocities well below this limit. We therefore consider any $v$sin$i$ measurement above 10 \kms~to be an anomalously rapid rotator. We find that none of the 9 stars discussed above have a $v$sin$i > 10$ \kms. This is not surprising since a rotational velocity of 10 \kms~corresponds to a $~4$-day rotation period for a turnoff star in the cluster. Given this detection limit, most of the stars in our sample would not have $v$sin$i$s above the WOCS velocity resolution limit regardless of inclination angle, and the rest would go undetected if the rotational axes of the systems are somewhat inclined. 

We do, however, detect two additional stars with $v$ sin $i$ $> 10$ \kms, WOCS 3001 and 11006. These stars have $v$sin$i$ measurements of 14.7 \kms and 18.1 \kms, respectively. For these stars we convert $v$sin$i$ to rotation period using the technique explained in \citet{Leiner2018}. Briefly, we fit photometric radii to the CMD position of each star. Using this radius we convert the observed rotational $v$sin$i$  to a distribution of periods assuming a random, uniform distribution of inclinations. We adopt the median value of this period distribution as the rotation period, and report the interquartile range of values as the uncertainty. Using this method, we derive rotation periods of  $2.6_{1.4}^{3.4}$  days for WOCS 11006 and  $3.3_{1.7}^{4.3}$ days for WOCS 3001. 

 We adopt these values in Table~\ref{table}, as these stars do not have well measured photometric rotation periods (we discuss this further in Section~\ref{section:stars}). Despite the lack of a photometric signal, we consider these reliable detections and include these two systems in our sample using these spectroscopic rotation periods.

\section{Discussion of Individual Stars}\label{section:stars}
In Table~\ref{table} we list the stellar and orbital properties of the rapidly rotating main-sequence stars. All these rapid rotators are high-probability proper-motion and radial-velocity members of M67. Nevertheless, there is a small probability any individual star in our sample may be a field star whose 3D motion overlaps with the motion of the cluster. However, the probability that more than one of these stars is a field contaminant is negligible \citep{Mathieu2003}.

Overall, 2 of 11 stars in our sample are selected based on spectroscopic $v$sin$i$'s, and 9 of 11 are selected from K2 light curves. Of these 9 stars, 5 are very secure detections of rapid rotation in which we find the same periods in Campaign 5 light curves from both the Oxford pipeline and CfA pipeline, and detect this period in CfA light curves from Campaign 16. 

4 of the 9 are less secure detections that have complex signals and/or are not observed in both K2 Campaigns. These signals are more open to different interpretations or have higher possibility of being spurious, but we still include them as possible detections. We discuss these in detail below.

\subsection{Spectroscopic Detections}
\subsubsection{WOCS 11006}
WOCS 11006 is a single-lined spectroscopic binary. We do not have a final orbital solution, so we cannot derive well-constrained orbital parameters. Our radial-velocity observations cover a time baseline of more than 10,000 days and indicate the binary system is long-period ($> 3500$ days, and perhaps as long as $10000$ days) with a very large orbital eccentricity. The $v$sin$i$ derived from the WOCS spectra is 18 \kms.

The Campaign 5 periodogram for this star shows several low-amplitude peaks, including one at 2.6 days and one at 8 days. A stronger peak at 8 days is measured in a nearby companion, and thus this periodic signal may be background contamination. The 2.6 day peak appears to originate with 11006. The Esselstein pipeline also measures a periodic signal of 2.6 days, though it is also below their secure-detection threshold. The C16 light curve also shows a weak signal at a similar period of 3 days. These signals are all low confidence and we would not include this star in our sample based on the periodogram alone. However, the $v$sin$i$ conversion suggests a rotation rate of $2.6_{1.4}^{3.4}$ days, providing additional evidence that the photometric rotation period is real. We adopt this 2.6 day rotation period for our analysis. 

\subsubsection{WOCS 3001}
WOCS 3001 is a circular, 128-day single-lined spectroscopic binary that is the bluest star in our sample and notably bluer than the rest of the main sequence (Figure~\ref{CMD}). Despite its color, it was not included in the \citet{Geller2015} sample of blue stragglers because it fell too close to the blue hook region to be confidently identified as part of the blue straggler population. The detection of this star's elevated rotation rate provides additional evidence that it is indeed a relative of the blue stragglers. 

The $v$sin$i= 14.7$ \kms~measurement indicates that WOCS 3001 is a rapid rotator. However, WOCS 3001 does not show a rotation signal at short periods in the Campaign 5 periodogram, and in C16 shows only a very weak signal (P = 3.3 days) that is not significant. The C5 light curve does suggest a 20-day rotation signal, although due to the 75-day time baseline of the K2 observations, the detection of such long-period variability is not secure \citep{Esselstein2018}. Due to its unusually hot temperature for the cluster, WOCS 3001 is near the divide between stars with convective envelopes and those with radiative envelopes. As a result, it is possible that the star could be a rapid rotator without significant evidence of spot modulation.  

The Esselstein approach measures a signal at 2.3 days in the C5 lightcurves that is below their secure-detection threshold, and does not detect the 20-day signal. This rotation period would be compatible with the $v$sin$i$ measurement. The C16 light curve shows the strongest periodicity at $P = 3.6$ days, but it is again a weak signal below their detection criteria. Given the low reliability of all the measured signals, we conclude there is no clear period evident in the K2 light curves. In the analysis that follows, we adopt a rotation period of 3.3 days, which we derive from the $v$sin$i$ measurement using the technique outlined in Section~\ref{section:rotationalevolution}.

\subsection{Secure Photometric Detections}
\subsubsection{WOCS 14020}
In Campaign 5, WOCS 14020 shows two strong peaks in its periodogram, one at 4.4 days and one 4.5 days. The Campaign 16 light curves show the same signals. 

A rotation period of 4.4 days is also detected using the Esselstein et al. pipeline.  The light curve (Figure \ref{LC}) also shows a clear beating pattern. Given the close spacing of these two peaks, we suggest the two periods are evidence for differential rotation on the star, with two star spots at slightly different latitudes moving in and out of phase with each other (e.g. \citealt{Reinhold2015}). 

The spectrum of this star is single lined, and the mass function sets a minimum companion mass of 0.15 \Msolar. The lack of flux from a companion in the spectrum combined with the very low mass minimum strongly suggests either a white dwarf or a M-dwarf secondary star. Interestingly, the system is detected in both the GALEX NUV and FUV filters (effective wavelengths of $\lambda_{eff}= 2267$ and 1516, respectively; \citealt{Martin2005}) , indicating the system has a large UV excess. We show the spectral energy distribution in Figure~\ref{fig:SED}. The GALEX FUV measurement supports the presence of a hot, faint companion, likely a white dwarf star. The FUV flux excess is consistent with a $\sim13,000$ K C/O white dwarf. This temperature implies an age of $\sim300$ Myr \citep{Tremblay2011}, in general agreement with the age implied from the rotation rate of $\sim200$ Myr.

Narrow-band UV photometry and/or UV spectroscopy are needed to confirm this detection and provide better temperature and age estimates. 

\begin{figure*}
 \vspace{1cm}
 \centering
\subfigure[P$_\mathrm{rot}$= 7.7 days.]{\includegraphics[width=.95\linewidth, angle=0]{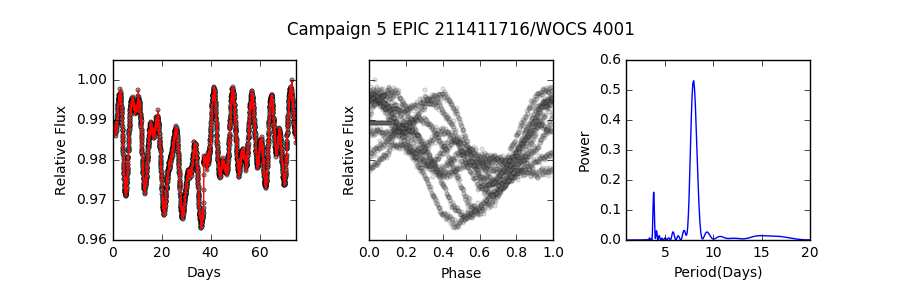}}
\subfigure[Two signals at P$_\mathrm{rot}$= 4.4 and 4.5 days, suggestive of differential rotation.]{\includegraphics[width=.95\linewidth, angle=0]{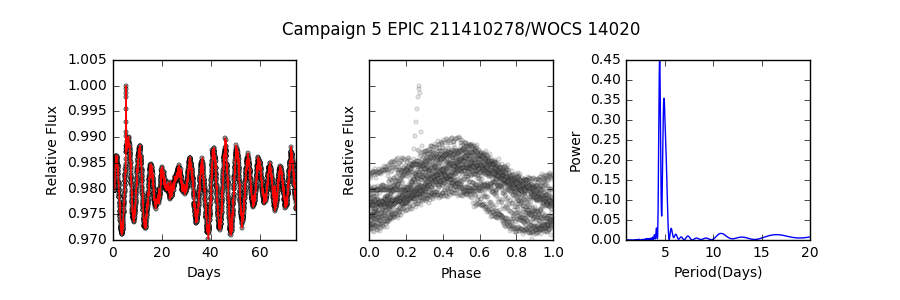}}
\subfigure[P$_\mathrm{rot}$= 3.7 days.]{\includegraphics[width=.95\linewidth, angle=0]{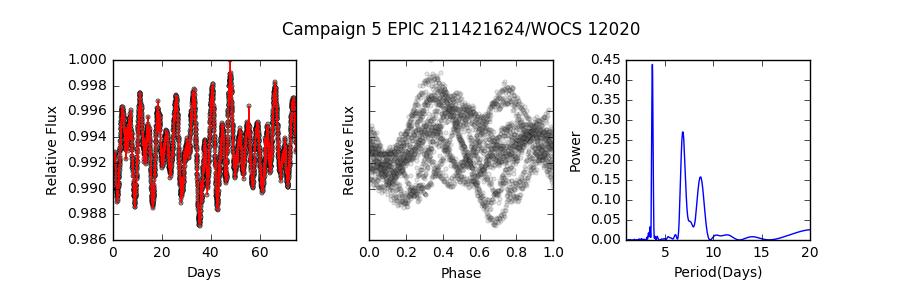}}
\end{figure*}

\begin{figure*}
 \vspace{1cm}
\centering
\subfigure[P$_\mathrm{rot}$=$20.5$ days in periodogram; also 3.3 days converted from $v$sin$i$. The phased light curve is folded on the 20.5 day period. ]{\includegraphics[width=.95\linewidth, angle=0]{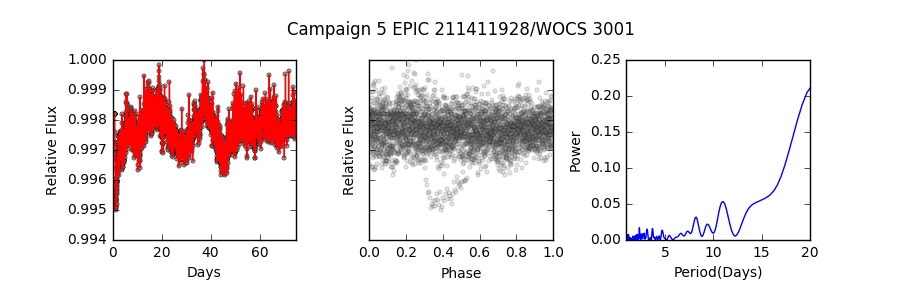}}
\subfigure[Marginal signal at P$_\mathrm{rot}$= 20.3; phased light curve shows an additional signal at 3.3 days.]{\includegraphics[width=.95\linewidth, angle=0]{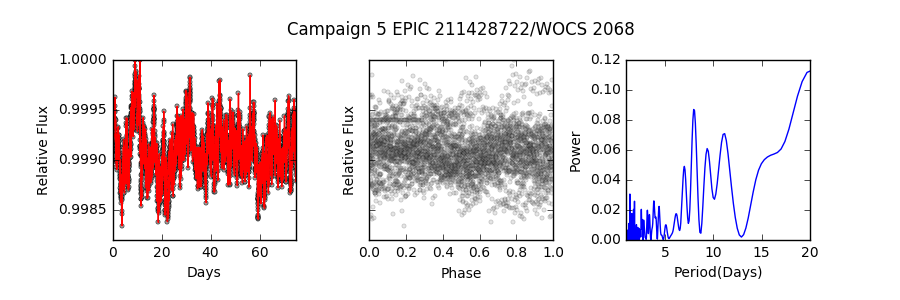}}
\subfigure[P$_\mathrm{rot}$= 4.5 days.]{\includegraphics[width=.95\linewidth, angle=0]{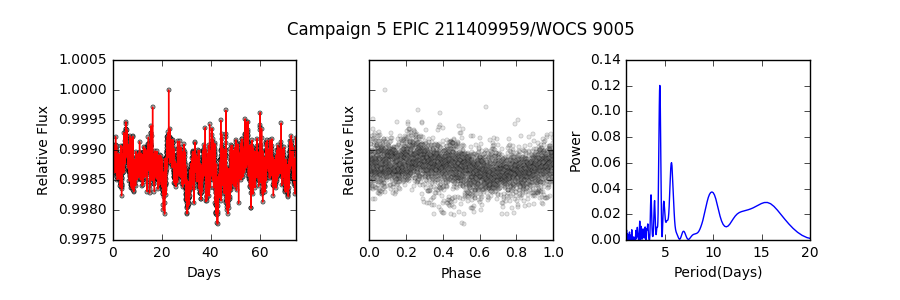}}
\end{figure*}

\begin{figure*}
 \vspace{1cm}
\centering
\subfigure[P$_\mathrm{rot}$= 2.3 days.]{\includegraphics[width=.95\linewidth]{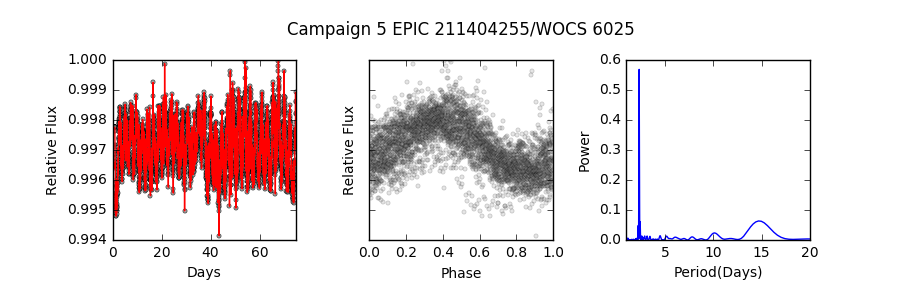}}
\subfigure[Weak signal at P$_\mathrm{rot}$= 2.6 days, but matches the rotation period converted from $v$sin$i$.]{\includegraphics[width=.95\linewidth, angle=0]{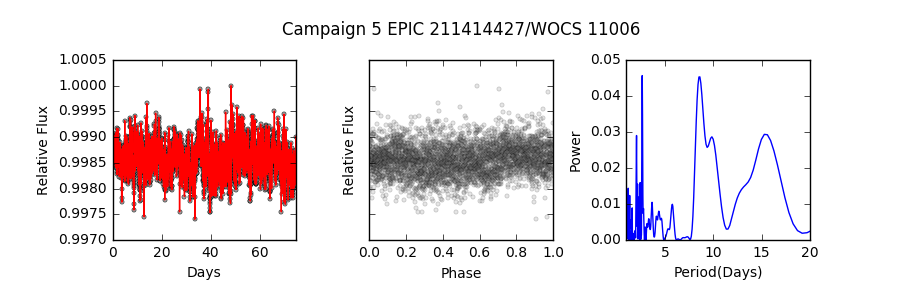}}
\subfigure[P$_\mathrm{rot}$= 5.6 days.]{\includegraphics[width=.95\linewidth, angle=0]{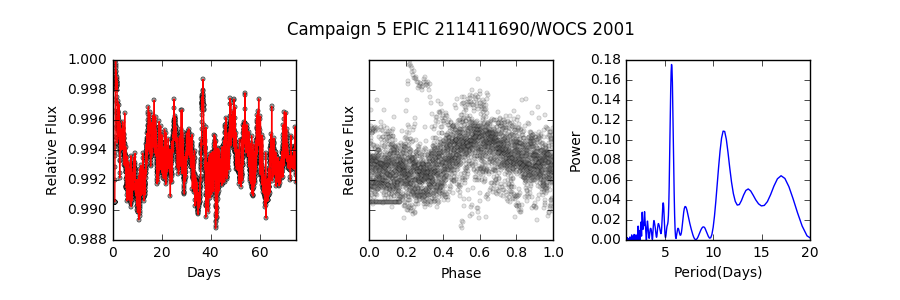}}
\end{figure*}

\begin{figure*}
\centering
\subfigure[Multiperiodic with the strongest peak at P$_\mathrm{rot}$= 5.9 days.]{\includegraphics[width=.95\linewidth, angle=0]{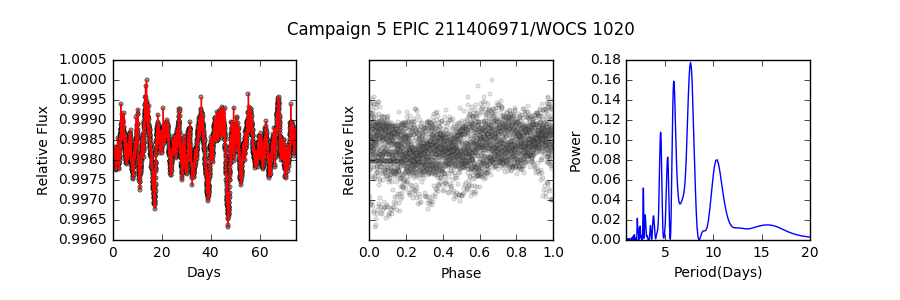}}
\subfigure[P$_\mathrm{rot}$= 8.0 days.]{\includegraphics[width=.95\linewidth, angle=0]{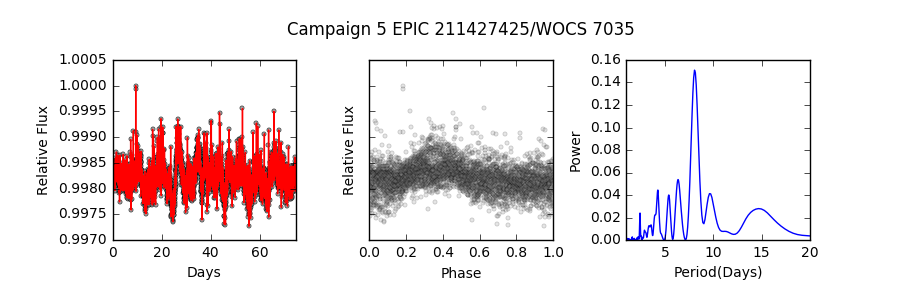}}
\end{figure*}

\begin{figure*}
 \vspace{1cm}
 \centering
\subfigure[P$_\mathrm{rot}$= 7.9 days as in C5.]{\includegraphics[width=.95\linewidth, angle=0]{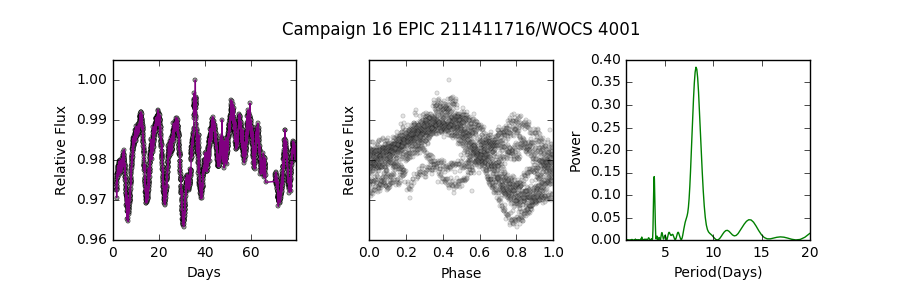}}
\subfigure[Two signals at P$_\mathrm{rot}$= 4.4 and 4.5 days, suggestive of differential rotation as in C5.]{\includegraphics[width=.95\linewidth, angle=0]{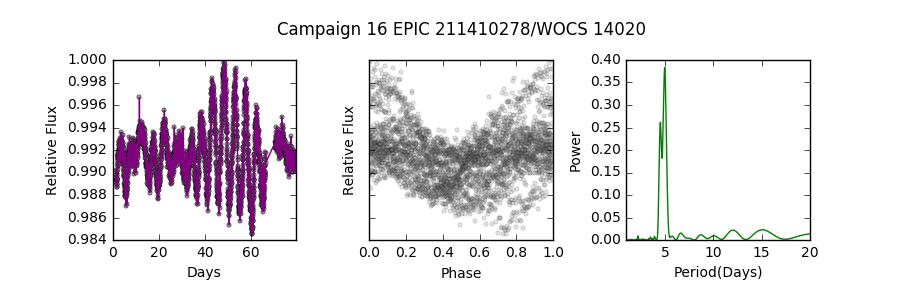}}
\subfigure[P$_\mathrm{rot}$= 7.6 days, twice the period of C5]{\includegraphics[width=.95\linewidth,angle=0]{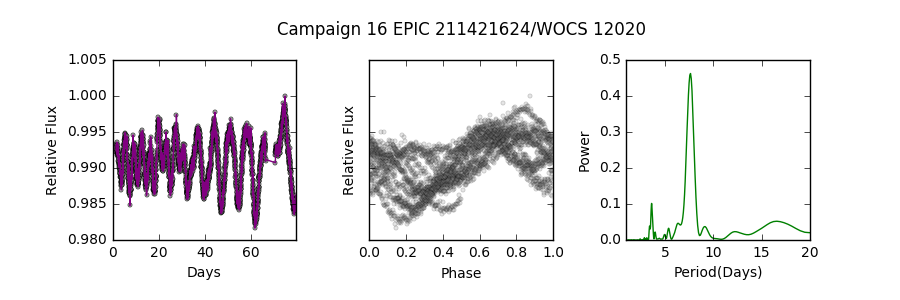}}
\end{figure*}

\begin{figure*}
 \vspace{1cm}
\centering
\subfigure[P$_\mathrm{rot}$=3.3 days converted from $v$sin$i$. The C16 light curve shows a very weak peak at $P= 3.6$, consistent with the spectroscopically derived rotation, which we use to fold the light curve. ]{\includegraphics[width=.95\linewidth, angle=0]{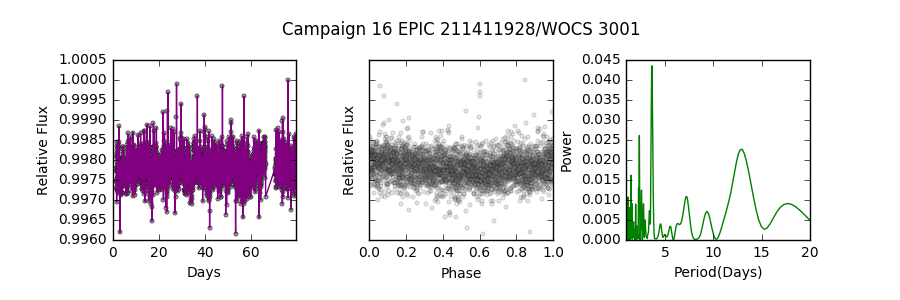}}
\subfigure[In C16 light curve shows variability at short periods as well as a marginal signal at longer periods (P= 15 day), similar to C5.]{\includegraphics[width=.95\linewidth, angle=0]{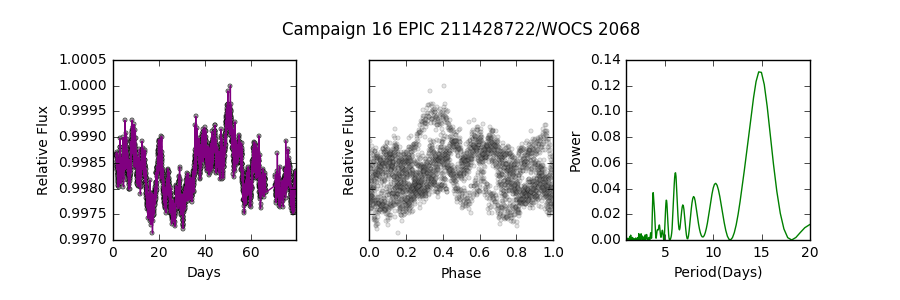}}
\subfigure[P$_\mathrm{rot}$= 4.5 days in C5. A much lower-amplitude signal around 5 days can be seen in the C16 power spectrum, but it is not clearly significant. Light curve is phased on the C5 period.]{\includegraphics[width=.95\linewidth, angle=0]{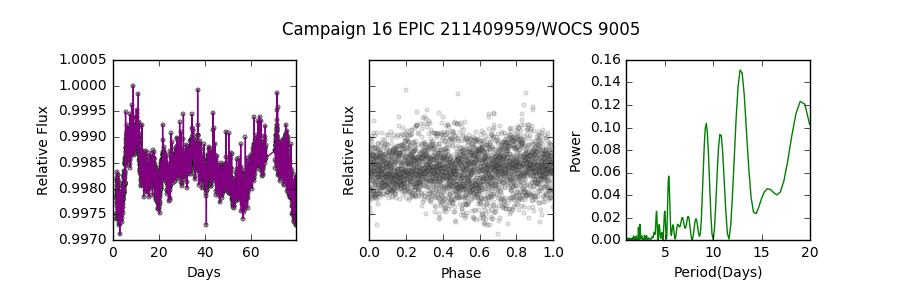}}
\end{figure*}

\begin{figure*}
 \vspace{1cm}
\centering
\subfigure[P$_\mathrm{rot}$= 2.3 days, as in C5.]{\includegraphics[width=.95\linewidth]{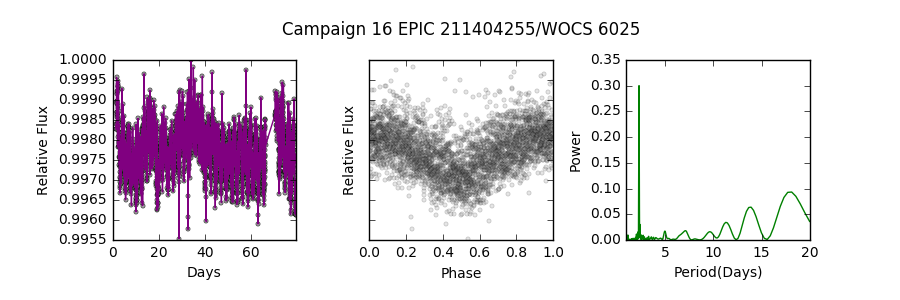}}
\subfigure[Low signal $P= 2.2$ days in C16, similar to C5.]{\includegraphics[width=.95\linewidth, angle=0]{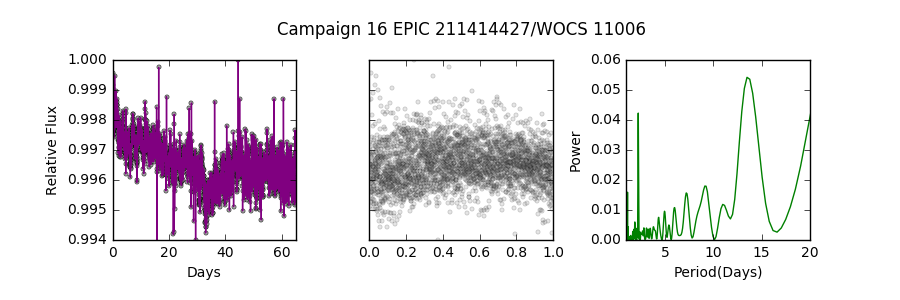}}
\subfigure[$P= 5.6$ days as in C16, same as C5. ]{\includegraphics[width=.95\linewidth,angle=0]{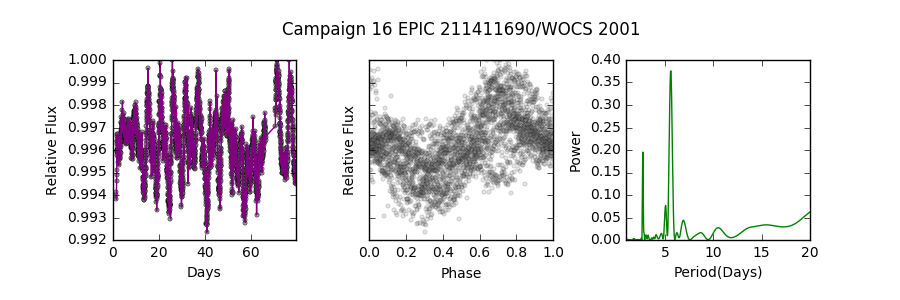}}
\end{figure*}

\begin{figure*}
\centering
\subfigure[C16 light curve shows lower amplitude peaks near 5 days]{\includegraphics[width=.95\linewidth, angle=0]{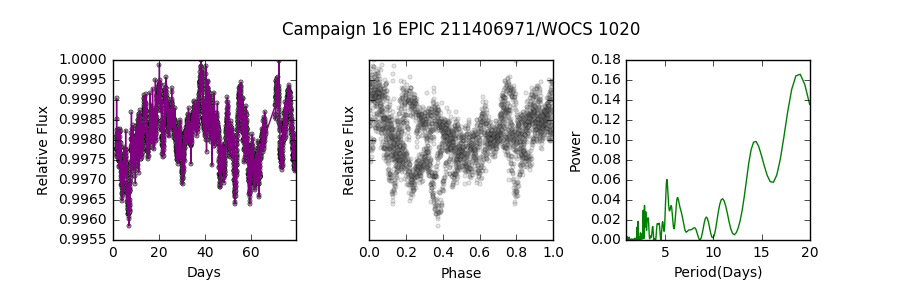}}
\subfigure[No Significant peaks in C16. P$_\mathrm{rot}$= 8.0 days in C5. Lightcurve is phased on the 8.0 day signal from C5.]{\includegraphics[width=.95\linewidth, angle=0]{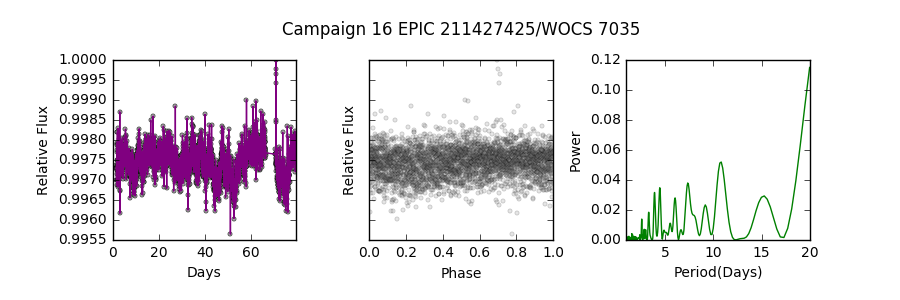}}
\caption{Light curves and periodograms of 11 main-sequence rapid rotators for C5 (red and blue) and C16 (purple and green).  On the left we show the full K2 Campaign 5 or r16 light curve for each star. In the middle we show the phased light curve folded on the dominant period, with the y-axis scaled as in the left plot. On the right, we show Lomb-Scargle periodogram for each star.}\label{LC}
\end{figure*}

\subsubsection{WOCS 4001, 6025, 12020}

These systems are all single-lined spectroscopic binaries. WOCS 6025, 4001, and 12020 have very clear signatures of spot modulation in their C5 and C16 light curves at levels of up to a few percent, and strong peaks in their CfA periodograms with periods consistent between C5 and C16 and that are independently confirmed by Esselstein.  

WOCS 4001 has a clear period of $P= 7.9$ days in both C5 and C16 light curves. Esselstein find a slightly longer period of $P= 8.0$ days. It is located in the cluster core, and thus has many bright neighbors that we examine carefully to determine if they may be contaminating the light curve. A  bright (Kp= 12.69) nearby star (WOCS 1001) does display variability at a similar period, but a comparison of the light curves reveals they have different shapes and are out of phase with each other.

For WOCS 12020, the Campaign 5 light curve shows the strongest peak at $P= 3.7$ days, with a lower peak at twice this period. In contrast, Campaigns 16 finds the strongest peak is P= 7.6 with the 3.7 day period the secondary peak. It can happen that a star has two spots on opposite sides, causing significant periodicity at half the true rotation period. We therefore interpret 7.6 days as the true period, which we report in Table \ref{table}.  

WOCS 6025 has a period of $P= 2.3$ in C5 and C16 light curves, and is also detected in Esselstein et. al light curves with the same period. 

All three stars are long-period (P$_\mathrm{orb} > 100$ day) binaries. All but WOCS 6025 have secondary mass limits consistent with white dwarf companions and thus with being candidate post-mass-transfer systems  (Figure~\ref{fig:wdperiod}). WOCS 6025 has a large secondary mass limit of $1.1$ \Msolar, more compatible with an F or G main-sequence star, though the spectra and the spectral energy distribution (SED) of the system do not reveal any evidence of such a companion. Another possibility could be that the system is a triple system composed of a near-turnoff primary star, and a secondary that is a close binary composed of two low mass stars ($\sim 0.5$ \Msolar). If these stars were tidally locked in a 2.3 day orbit, this could also explain the origin of the periodic signal. 

\subsubsection{WOCS 2001}
WOCS 2001 shows strong single peaks at P = 5.6 days in both of the C5 and C16 periodograms. This period is also confirmed by Esselstein measurements. WOCS spectra indicate the star is not a velocity variable and therefore we classify it as single. If this star has a binary companion that is not spectroscopically detected, it must be wide ($P_\text{orb} \gtrsim 10000$ days) or viewed very close to face on (although assuming rotation and orbital axes are aligned, this would make detecting a photometric rotation period unlikely).  

Assuming it is a true single star, this system would not have formed from mass transfer or a Kozai-induced merger in a triple system, as both scenarios are expected to leave behind binary systems. A merger of a close main-sequence binary system through internal processes such as magnetic braking (e.g. \citealt{Andronov2006}), or a dynamical scenario may also be a plausible origin. Dynamical collisions often leave the collision product bound in a binary or higher order system. However, the collision product may also be left as an unbound single star like WOCS 2001, although this outcome is less probable \citep{Fregeau2004}. 

\subsection{Possible Photometric Detections}
\subsubsection{WOCS 2068}\label{section:2068}
WOCS 2068 is a long-period (8567 days), highly eccentric (e= 0.859) binary. The SED of the system suggests the binary consists of two stars (see Figure~\ref{fig:SED} and Section~\ref{section:SED}). 

One of these stars has an unusually hot temperature for the cluster, placing it slightly to the blue of the main-sequence turnoff and suggesting it is a blue straggler. The other is located on the cluster turnoff.

This star is multi-periodic, showing a marginal period detection of 15-20 days, as well multiple peaks at shorter periods indicating variability on a timescale of several days. 

\begin{figure*}[hbpt]
\centering
\includegraphics[width= .9\linewidth]{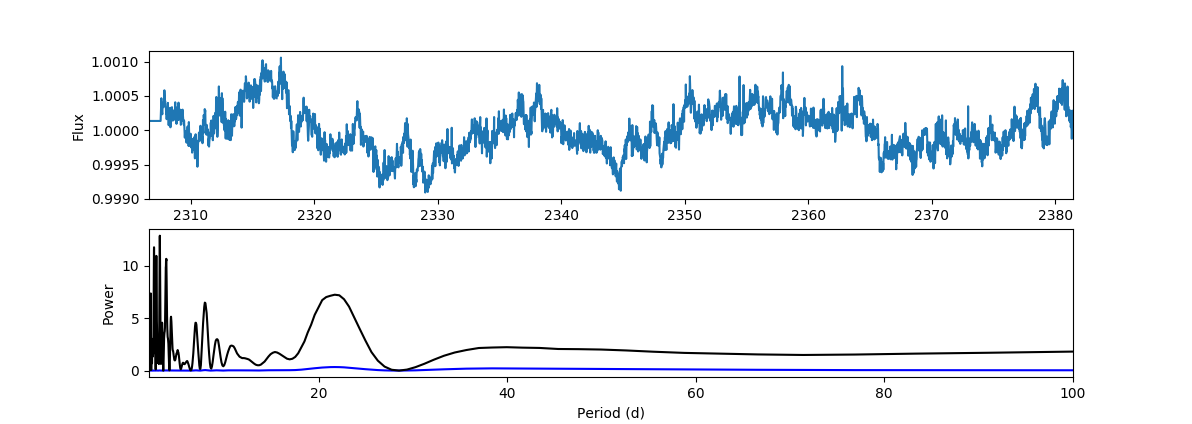}
\caption{The light curve (top) and power spectrum (bottom) of WOCS 2068 from the pipeline of Esselstein et al. 2018. In the power spectrum, the blue line shows the raw power and the black line shows the normalized power as defined in Esselstein et al 2018. The normalized power spectrum shows the strongest power at $\sim3$ days, as well as a signal at 20 days, though the multi-periodic nature of the star and relatively low amplitude put it below their formal detection criteria. }\label{fig:2068}

\end{figure*}

In Figure 3, we have folded the light curve on the 20-day-period signal. The phase-folded light curve shows additional variability on shorter period timescales ($\sim3$ days). The Esselstein pipeline also measures a $\sim3$-day and 20-day rotation period, though the multi-periodic nature of the star and the relatively low amplitude put it below their detection criteria. We show the light curve and power spectrum from the Esselstein analysis in Figure~\ref{fig:2068}. The same type of variability is observed in Campaign 16 light curves (Figure ~\ref{LC}). We note that although the 15-20 day period technically shows more power in Figures ~\ref{LC}, longer period signals are much more likely to be spurious due to the K2 instrumental systematics. Detections of short period variability, even at lower power, are more reliable (see \citealt{Esselstein2018} for a detailed discussion).

We consider this a strong detection of short-period variability, but due to the complexity of the light curve and the low amplitude of the variability we cannot measure a precise period nor definitively attribute the variation to rotation. The short-period signals in the light curve may be consistent with small spots on a rotating star, but they could also be consistent with stellar pulsation. Normal turnoff stars in M67 are not in the instability strip, nor observed to be pulsating. However, the unusually hot 6800 K companion inferred from the SED fit (Section~\ref{section:SED}) would be near the red edge of the $\gamma$ Doradus region \citep{Handler1999, Balona2018}, and it can be difficult to differentiate between rotation and $\gamma$ Doradus pulsations \citep{Balona2011, Rebull2016}. However, $\gamma$ Doradus pulsators usually have periods of 0.4-3.0 days \citep{Kaye1999}, somewhat shorter than the timescale of the variability of this source. 

Regardless of the interpretation of the light curve, it is likely that the system has been through a stellar interaction of some kind given that the SED indicates one component is hotter than the cluster main sequence turn off and is either a rapid rotator or a pulsator, neither of which would be expected for a normal main sequence star in the cluster. 

If we attribute the observed periodicity to be rotation, it suggests the system is composed of one slowly rotating main-sequence star with period of 15-20 days, typical for the cluster, though we stress that this longer period may be due to K2 systematics rather than true stellar variability. The other star in the system has a short rotation period of $\sim3$ days. The system is not a main sequence-white dwarf binary, the expected outcome mass transfer formation. It is unlikely a white dwarf would have been ejected from this system in a dynamical encounter given the expected encounter rates in M67 and the young age of the system inferred from its rapid rotation \citep{Leigh2011}. Mass transfer is therefore not a likely explanation for this system. Instead it may be a merger or dynamical collision. In such a merger scenario, the initial system may have been a hierarchical triple consisting of a short-period inner binary and a distant triple companion. The inner binary was then driven to a merger, either by magnetic braking or through Kozai-Lidov oscillations, with the former tertiary now the observed wide secondary.

Alternatively, the system may have resulted from a dynamical encounter involving at least one binary star in which two stars collided (e.g. \citealt{Leonard1989,Sills2001}). Scattering experiments show that binary systems resulting from dynamical collisions tend to result in very high eccentricity orbits \citep{Fregeau2004}, as is seen in this system. In contrast, a  merger of an inner binary in a triple system does not favor any particular eccentricity \citep{Naoz2014}, and mass-transfer origin preferences low eccentricity outcomes (e.g. Figure~\ref{fig:elogp}). 

Additionally, this binary is located in the cluster halo, not in the core as is expected for most binaries due to mass segregation. Dynamical encounters can impart a recoil velocity to a star (e.g. \citealt{Phinney1991}), pushing the orbit farther into the halo or perhaps ejecting it entirely, which could explain this binary's less probable location within the cluster.

Taking this information together, the properties of WOCS 2068 may fit best with a recent stellar dynamical encounter that resulted in the collision of two main-sequence stars. A merger in a triple system is also possible, but a mass-transfer origin is not consistent with the observed system.

\subsubsection{WOCS 9005}
WOCS 9005 is a long-period binary system showing periodicity in the C5 lightcurve at $P= 4.5$ days. The same period is detected using the pipeline of Esselstein et al., but the amplitude is below their secure-detection threshold. Our visual inspection confirms a single, clean periodic signal in the periodogram and visible variability in both phased and unphased light curves matching this period. However, the C16 light curves do not show strong periodicity at this or any other period. There is a slight peak near $5$ days, and visual examination of the light curve suggests a low amplitude $\sim5$ day signal may be visible at the beginning of the C16 observations, which disappears over the course of the observations. These features may hint at the presence of a rotation signal in C16, but they are not conclusive.  

It is possible that small spot(s) visible on the star during C5 were weaker or no longer present during the C16. It is also possible that the signal visible in C5 was spurious. Given the strong detection in C5, we include 9005 in our sample, but caution that the detected period may be unreliable.

\subsubsection{WOCS 7035 and 1020}
Both WOCS 1020 and WOCS 7035 are non-velocity variable. If these stars do have companions that have avoided detection, these systems are either wide ($P_\text{orb} \gtrsim 10000$ days), or viewed close to face on.

The C5 lightcurve of WOCS 7035 shows a single peak in its periodogram at $8.0$ days, in agreement with the ACF period of Esselstein et al. However, we observe no strong periodicity in the C16 light curves. As with 9005, we include 7005 in our sample with caution.

WOCS 1020 shows several closely spaced peaks in the C5 periodogram, with the strongest at 5.9 days. Esselstein et al. do detect variability in this star with a similar period, but do not classify it as a significant detection due to the multi-periodic nature. The C16 light curve again show multiperiodic variability, but with a lower amplitude signal around 5 days. It also shows a longer period signal around 20 days, but such long period signals are not very reliable in K2 data. \citet{Esselstein2018} introduce a process to normalize periodograms to remove background signals at long periods, which we have not done here, and given the total lack of detection of a long period signal in C5 in both pipelines it is not clear if this signal is significant. 

While this star is clearly variable, the nature of the variability is not certain. The light curve could be consistent with multiple small starspots, but as with WOCS 2068 this star has an SED temperature that places it in the $\gamma$ Doradus region (Section~\ref{section:SED}). However, the 5.9 day period is more consistent with a rotation signal given the typical 0.4-3.0 periods of $\gamma$ Doradus variables \citep{Kaye1999}. We therefore classify this star as a probable rapid rotator.

If these stars are both rapidly rotating single stars, they would not have formed from mass-transfer or Kozai-induced mergers in triple systems, as both scenarios are expected to leave behind binary systems. Instead, they may have formed through dynamical collisions between main-sequence stars as we suggest for WOCS 2001.  

Another possibility is a merger of a close main-sequence binary system through internal processes such as magnetic braking (e.g. \citealt{Andronov2006}). 

\section{Characteristics of Rapid Rotators}

 In Figure ~\ref{CMD} we mark the location of these 11 stars in a color-magnitude diagram of M67 members. For reference, we show the rotation periods of these stars compared to standard stars in M67 \citep{Barnes2016} and those predicted by gyrochronology relations of \citet{Angus2015}(Figure~\ref{fig:gyro}). We discuss the properties of this sample below. 

\subsection{Binary Fraction}
Eight of the 11 systems in our sample are binaries, for a binary fraction of 73\% $\pm 31$\%. For comparison, the spectroscopic binary fractions ($P_\mathrm{orb} < 10^4$ days) of M67 and other old open clusters are observed to be in the range of 20-30\% \citep{Geller2015, Milliman2014, Geller2009}. These rapid rotators thus have about 3 times the binary fraction expected for a typical main-sequence population.

Classical blue straggler populations in old open clusters are observed to have similarly high binary fractions. In M67 itself, the blue straggler binary fraction is 80\% \citep{Geller2015}. In the 7-Gyr open cluster NGC 188, 76\% $\pm 19$\% of blue stragglers are observed to be spectroscopic binaries within a similar period domain \citep{Mathieu2009}.  Thus the observed high binary fraction among these rapid rotators is consistent with our hypothesis that they are lower-luminosity analogs of the blue stragglers formed through similar binary evolution channels.

\subsection{Orbital Properties}

\begin{figure}
\centering
\vspace{.5cm}
\includegraphics[angle=0, width= .9\linewidth ]{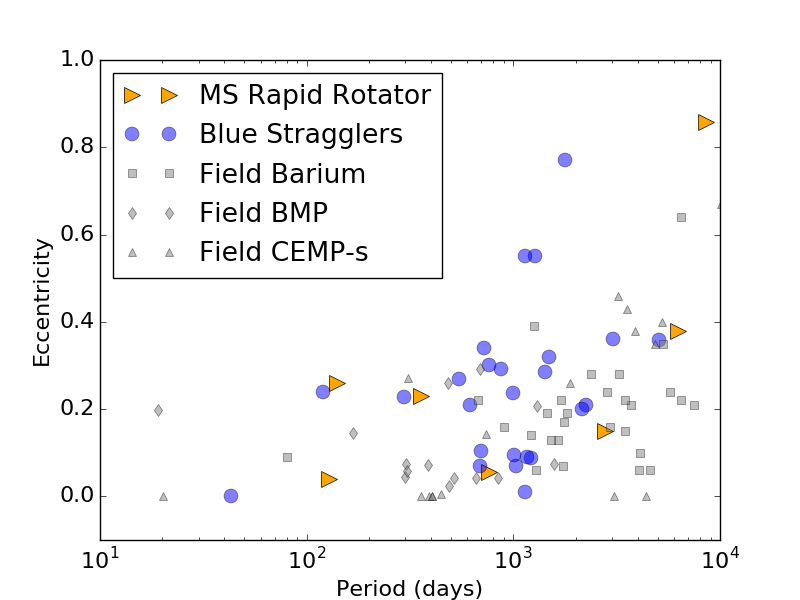}
\caption{A comparison of the distributions of periods and eccentricities of the M67 main-sequence rapid-rotators (orange triangles) to the the binary blue straggler populations of several old open clusters (blue circles) and post-mass-transfer binaries observed in field (gray symbols), including carbon-enhanced metal poor stars with s-process enrichment \citep{Jorissen2016, Hansen2016}, blue metal-poor stars \citep{Carney2005}, and barium stars \citep{Jorissen1998}. We do not include WOCS 11006 on this diagram, as we do not have a complete orbital solution, but it appears to be a long-period, high-eccentricity system that would fall in the upper right corner of this diagram. } \label{fig:elogp}
\end{figure}
 
 We show the eccentricity-period distribution of the 8 binaries in our sample in Figure~\ref{fig:elogp}. For comparison, we also show the eccentricity-period distribution of field barium stars, CEMP-s stars, blue metal poor stars, and blue stragglers in 3 old open clusters NGC 188 \citep{Geller2009}, M67 (Geller et al. 2019, in prep), and NGC 6819 \citep{Milliman2014}. Most of the binaries in our sample of rapid rotators have low eccentricities that fall within the eccentricity-period locus of these post-mass-transfer binaries. 

Typical populations of long-period solar-like main-sequence binaries have  eccentricities in a Gaussian-like distribution about a mean of e= 0.4 (e.g. \citealt{Meibom2006, Raghavan2010}). On the other hand, post-mass-transfer binaries show lower eccentricities, presumably because these systems go through substantial tidal dissipation before the onset of Roche lobe overflow. Even though blue stragglers and the other post-mass-transfer binaries often do not have circularized orbits, they do show lower eccentricities than solar-type main-sequence binaries at orbital periods of $\sim1000$ days \citep{Mathieu2009, Jorissen1998, Hansen2016, Jorissen2016, Carney2005}. Some do have circular orbits, as do two binaries in our sample of rapid rotators (WOCS 3001 and WOCS 12020), suggestive that both of these systems have been through mass transfer. 

On the other hand, two of the 8 binaries, WOCS 11006 and 2068, have much larger eccentricities and longer periods than typical. Such large eccentricities are perhaps more compatible with dynamical formation, as we suggest for WOCS 2068 (Section~\ref{section:2068}). 
 
We note that 3 of the binaries in our sample (WOCS 3001, 4001, 14020) have orbital periods of just a few hundred days, shorter than all but three of the observed blue stragglers in NGC 188, M67, and NGC 6819. If these stars are indeed post-mass-transfer binaries, their orbital periods suggest they result from Case B mass transfer (mass transfer from an RGB donor). These 3 short-period systems resemble WOCS 5379 in the cluster NGC 188, a 120-day blue straggler-white dwarf binary. Gosnell et al. (2019, submitted) measure a precise white dwarf mass for WOCS 5379 and demonstrate it is a Case B mass transfer product. However, this observation is difficult to resolve with mass transfer theory because a wide range of models and assumptions predict this system should have evolved to a common envelope. Like WOCS 5379, these three short-period systems in M67 may be interesting probes of the shortcomings of mass-transfer theory and the criteria for common envelope evolution.  
Additionally, it is interesting that more of these short-period systems show up in this lower luminosity domain than among the blue straggler population. This could hint that they form from binary systems with initially lower-mass secondaries, or that they do not accrete as much mass from their companions and are indicative of more inefficient mass transfer than the blue stragglers. These three systems are excellent candidates to model in more detail, as their evolutionary pathways may help constrain these uncertain aspects of mass-transfer physics.

\subsection{Companion Masses} 
\begin{figure}
\centering
\vspace{.5cm}
\includegraphics[angle=0, width= .9\linewidth ]{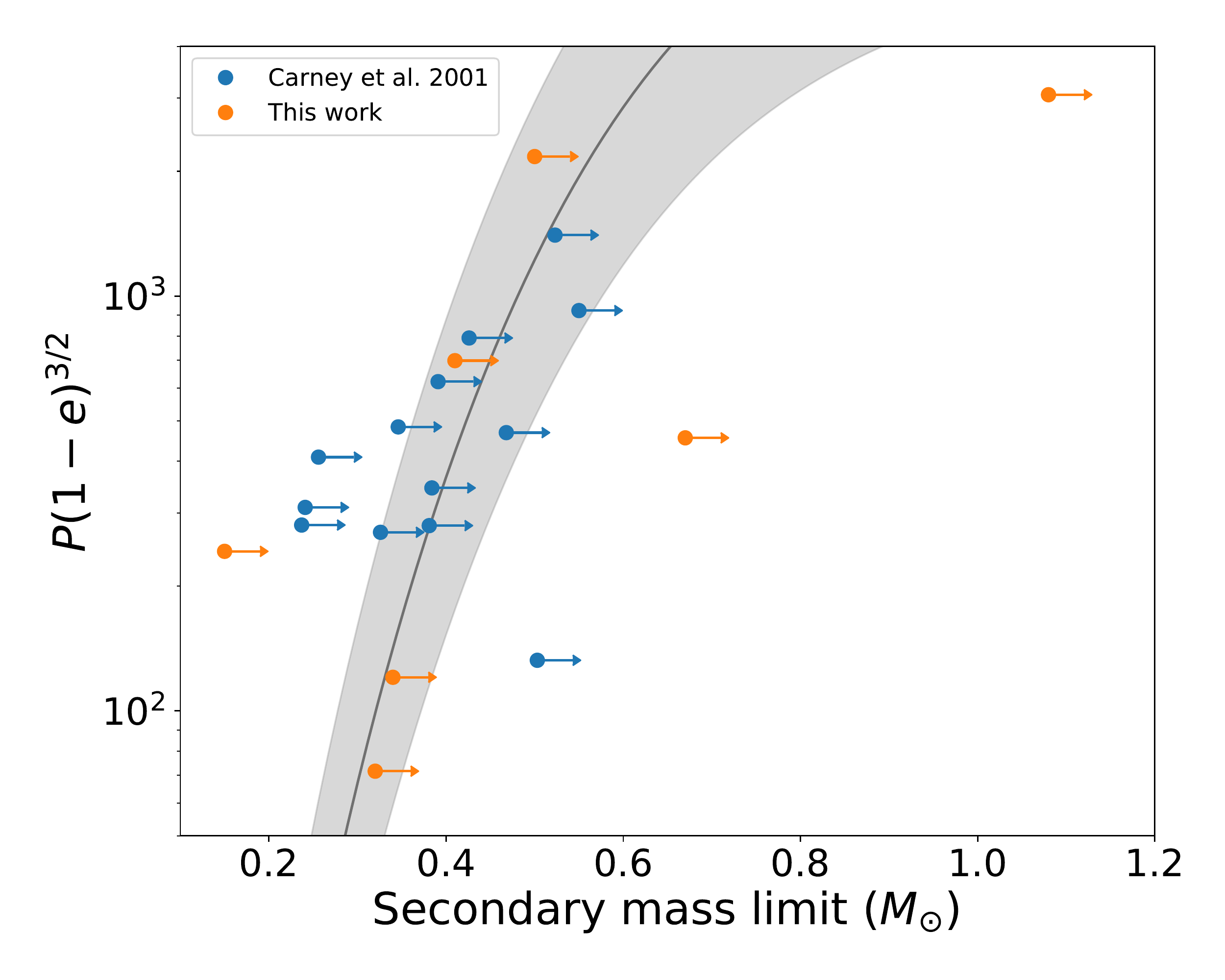}\label{WDPeriod}
\caption{We show the relationship between orbital separation at periastron and secondary mass for a sample of field blue straggler binaries from \citet{Carney2001} (blue points) compared to our sample of rapidly rotating main-sequence binaries (orange points). We also show in gray the \citet{Rappaport1995} theoretical period-white dwarf mass relationship for binaries resulting from stable mass transfer. We note that the comparison to the Rappaport relationship is only true for post-mass transfer systems with white dwarf companions. All points show lower limits on the secondary masses derived from the binary mass functions. WOCS 11006 is not included in the plot, as we have no secure orbital solution.}\label{fig:wdperiod}
\end{figure}

For 7 of the 8 binaries in our sample, we have orbit solutions that enable us to determine a binary mass function, f(m). From this function, we can derive lower limits on the mass of the secondary after adopting a mass for the primary. We do this by fitting a stellar evolutionary track to the CMD position of each system, recognizing that standard stellar evolution theory may not be accurate for these stars. We find that the primary stars range in mass from 0.9-1.35 \Msolar. Using these primary masses, we derive the minimum mass for each secondary star.  These secondary mass lower limits are shown in Figure~\ref{fig:wdperiod}, plotted against periastron separation. 

We find that 5 of the 7 binaries fall well within the period-secondary mass range expected for post-mass-transfer white dwarf-main sequence binaries \citep{Rappaport1995}. Two systems, WOCS 2068 and WOCS 6025, appear to have secondaries more massive than expected if their companions are white dwarfs. This is as expected for WOCS 2068, as the SED indicates contributions from two main sequence stars (Section ~\ref{section:SED}. WOCS 6025 also has a substantially more massive companion ($> 1.1$ \Msolar) than expected for a white dwarf. The white dwarf initial-final mass relation predicts such a massive white dwarf would form from a very massive progenitor ($> 6$ \Msolar; \citealt{Kalirai2008}), far above the 1.3 \Msolar~turnoff of M67.  This companion is therefore more compatible with a main-sequence star. Another possibility could be that the system is a triple system composed of a near turnoff primary star, and a secondary that is a close binary composed of two low mass stars ($\sim 0.5$ \Msolar). If these stars were spotted and tidally locked in a 2.3 day orbit, this could also explain the origin of the periodic signal. However, we note that some known white dwarf-main sequence binaries do not fall on the expected \citet{Rappaport1995} relation (e.g. \citealt{Kawahara2018}, Gosnell et al. 2019, in prep). On their own, the secondary masses cannot definitely confirm or rule out the existence of white dwarf companions for these sources.

\subsection{UV Excesses Indicative of White Dwarf Companions}\label{section:SED}

To look for evidence of companions to our systems, we examine their spectral energy distributions (SEDs). We include UV photometry from GALEX \citep{Martin2005}, IR photometry from 2MASS \citep{Skrutskie2007} and WISE \citep{Wright2010}, and optical photometry from \citet{Geller2015}, originally obtained by \citet{Montgomery1993}. Our SED fitting routine is described in \citet{Leiner2016}. Briefly, we fit a grid of \citet{CastelliKurucz} models of varying temperature and radius using $\chi^2$ minimization. We fix the surface gravity to a typical main-sequence value (log $g= 4.0$), assume solar-metallicity models, and fix the reddening to the cluster value (E(B-V)= 0.041; \citealt{Taylor2007}). We note that the SED fits are not particularly sensitive to our choice for surface gravity, metallicity, or reddening within a range of reasonable values. We set the distance to 850 pc, the median of the 800-900 pc range found to M67 in the literature \citep{Geller2015}.

\begin{figure*}
\begin{center}
\subfigure[WOCS 4001]{\includegraphics[width=.3\linewidth, angle=0]{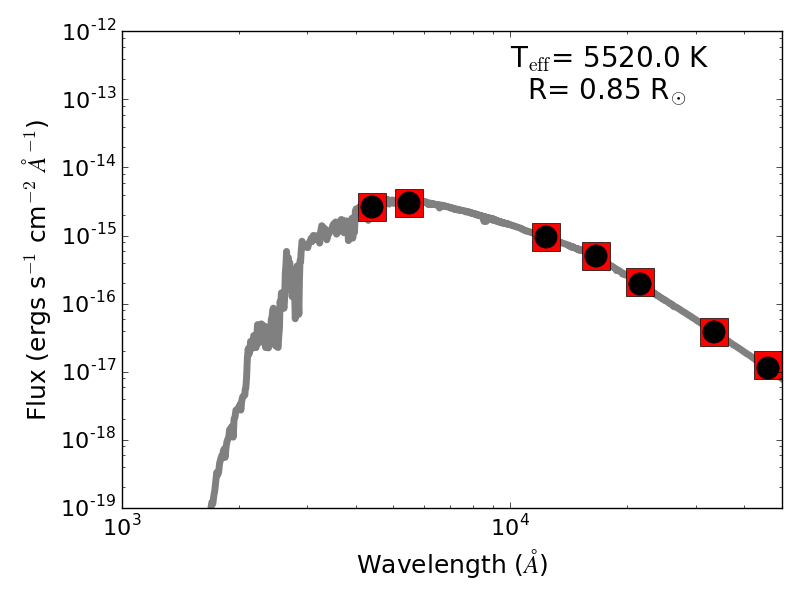}}
\subfigure[WOCS 14020]{\includegraphics[width=.3\linewidth, angle=0]{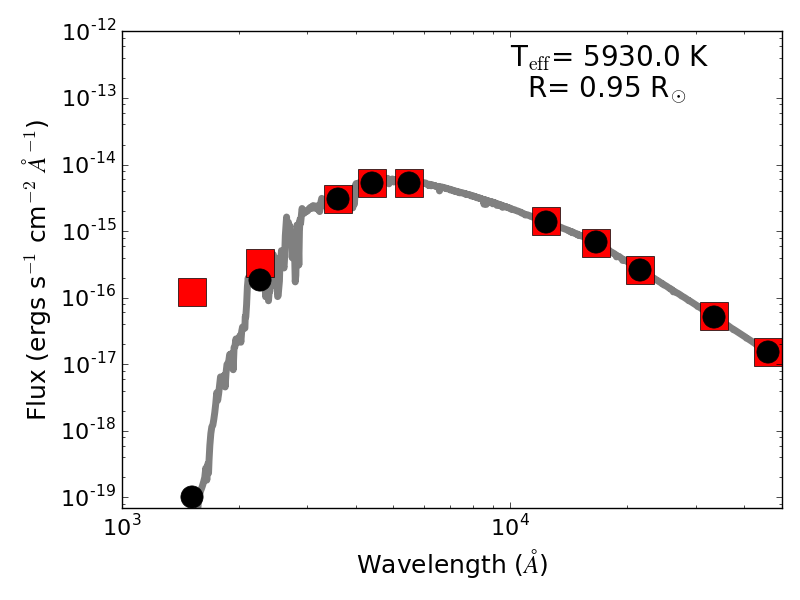}}
\subfigure[WOCS 12020]{\includegraphics[width=.3\linewidth, angle=0]{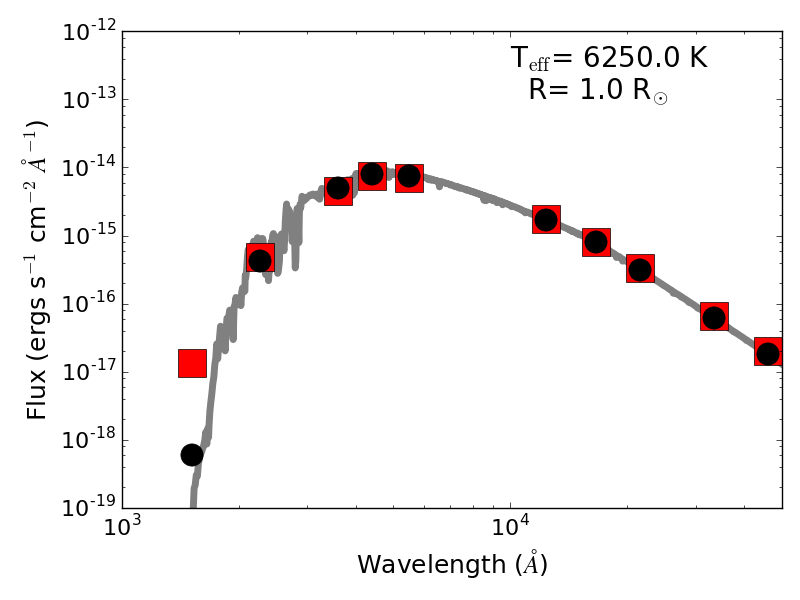}}
\subfigure[WOCS 3001]{\includegraphics[width=.3\linewidth, angle=0]{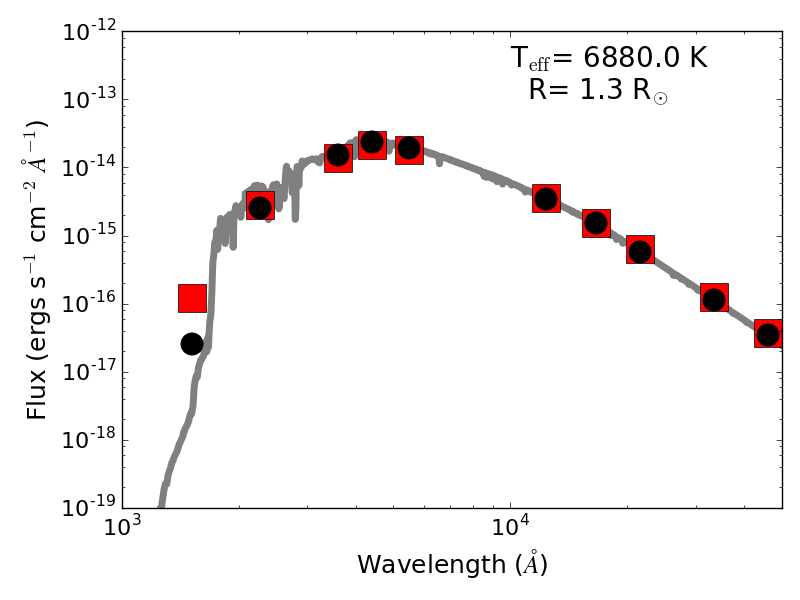}}
\subfigure[WOCS 2068]{\includegraphics[width=.3\linewidth, angle=0]{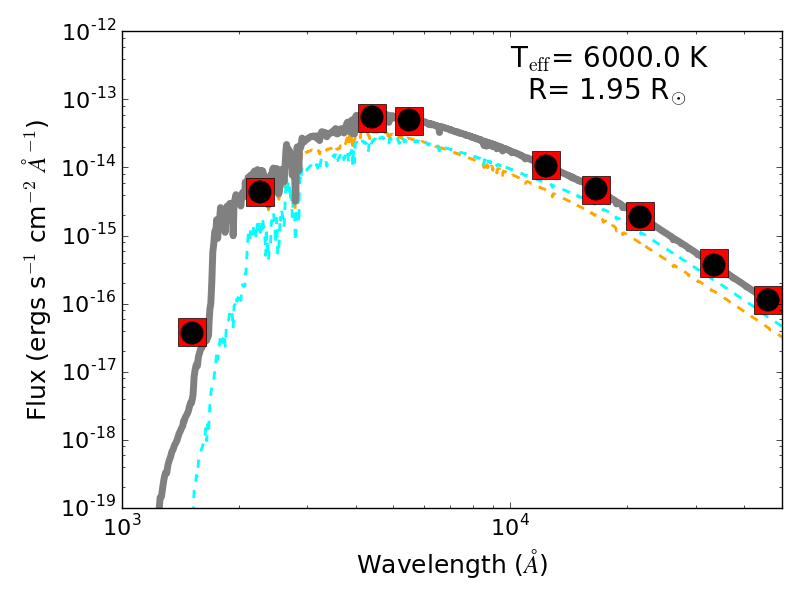}}
\subfigure[WOCS 9005]{\includegraphics[width=.3\linewidth, angle=0]{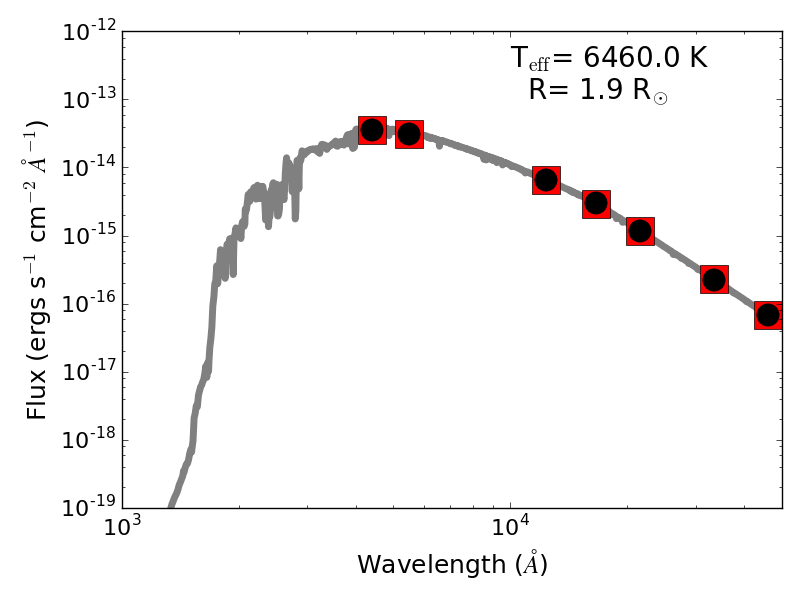}}
\subfigure[WOCS 6025]{\includegraphics[width=.3\linewidth, angle=0]{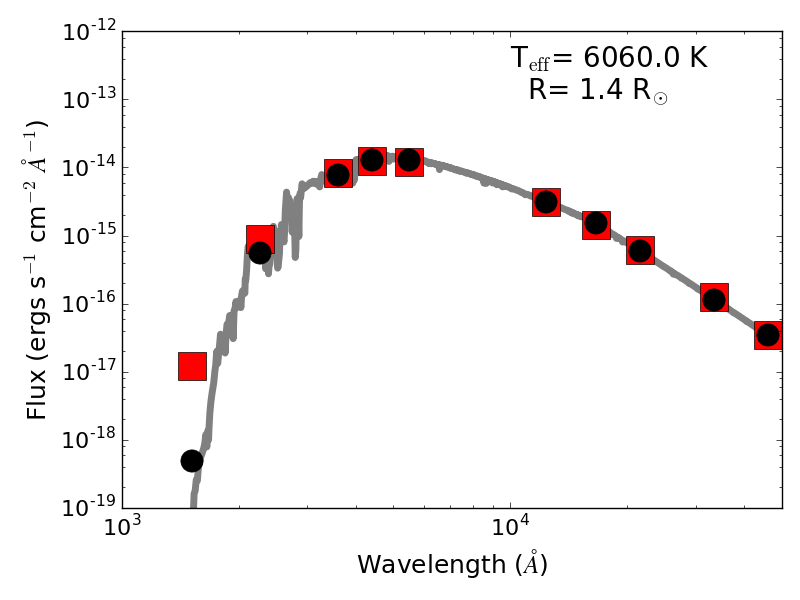}}
\subfigure[WOCS 11006]{\includegraphics[width=.3\linewidth, angle=0]{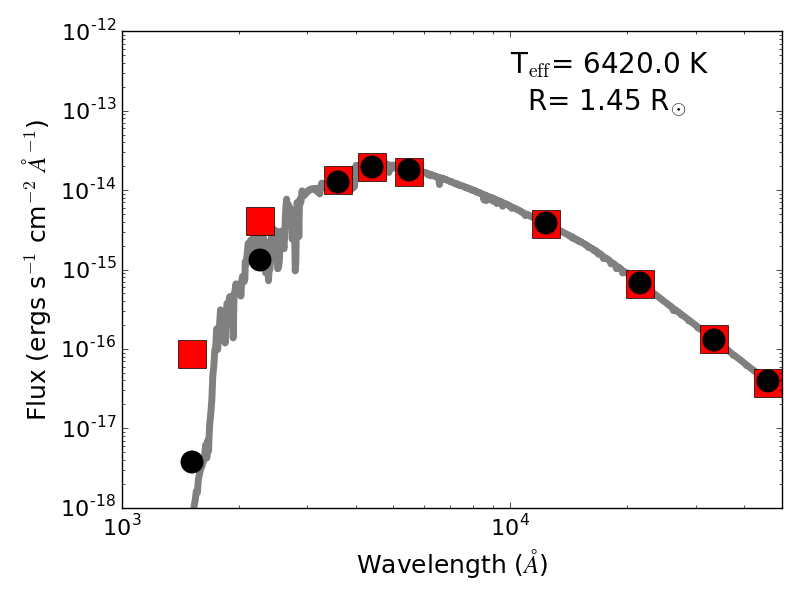}}
\subfigure[WOCS 1020]{\includegraphics[width=.3\linewidth, angle=0]{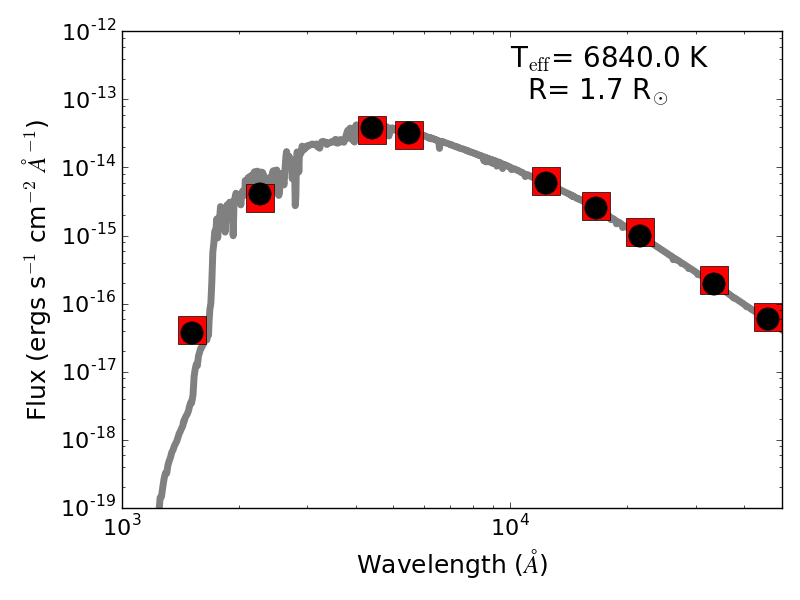}}
\subfigure[WOCS 7035]{\includegraphics[width=.3\linewidth, angle=0]{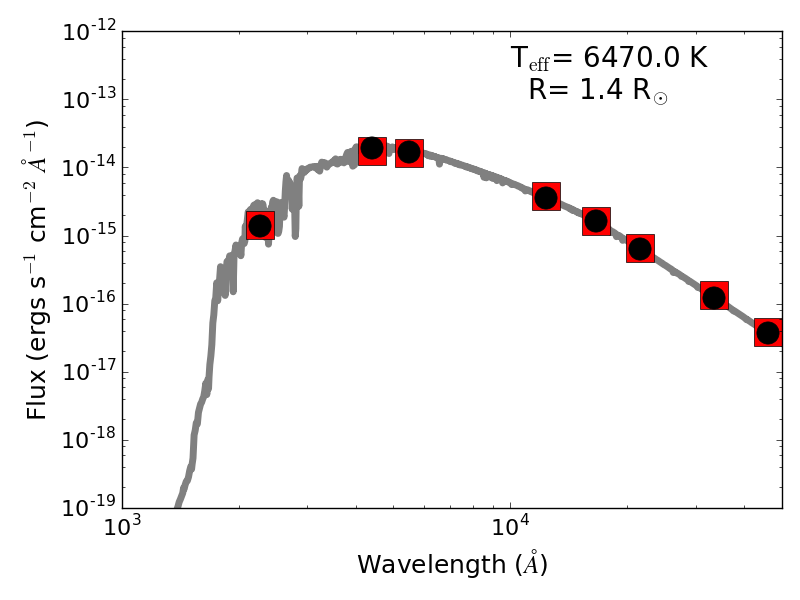}}
\subfigure[WOCS 2001]{\includegraphics[width=.3\linewidth, angle=0]{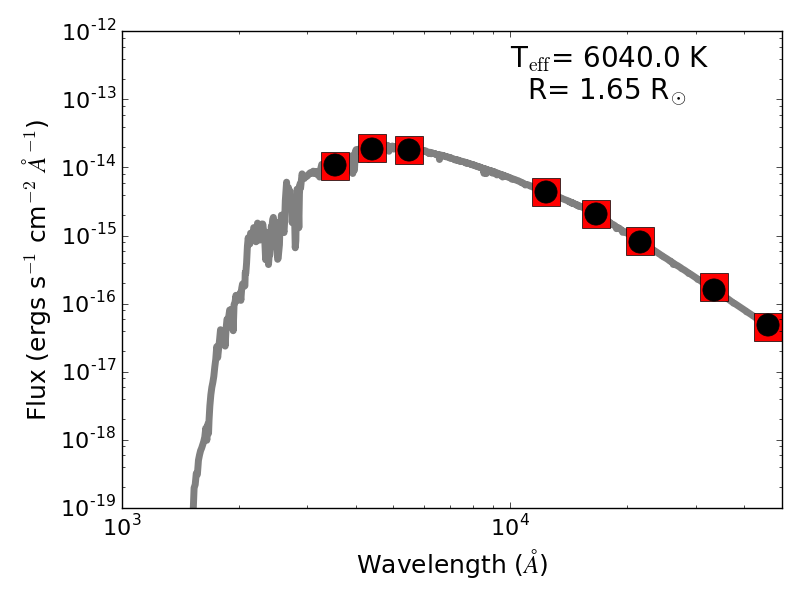}}
\end{center}
\caption{SEDs of the 11 rapid rotators in M67. We show the best-fit \citet{CastelliKurucz} spectrum in gray. Black circles are synthetic observations created by convolving the spectrum with filter transmission functions for 2MASS, WISE, GALEX, and Johnson UBV filters. The real observations are shown with red squares. For one system, WOCS 2068, we use a two star model. Here we plot the primary model (6000 K, 1.95 \Rsolar) in cyan, the secondary model (6800 K, 1.5 \Rsolar) in orange, and the combined flux in black.}\label{fig:SED}
\end{figure*}

We fit SEDs to these stars in three steps: 1) We fit a single, main-sequence model. 2) We fit a combination of 2 main-sequence stars. 3) We fit the flux excluding the GALEX FUV photometry, as this is the only bandpass that would have substantial flux from a hot white dwarf companion. If the best fit comes from Method 1, we characterize the star as being single or having a low-luminosity companion. If the best fit comes from Method 2, the system has a relatively bright main-sequence secondary. If the best fit comes from Method 3, this is indicative of a UV excess not well described by any main-sequence companion. These stars may have hot white dwarf companions contributing to their UV flux, or may have UV flux enhancements due to stellar activity. We show the best fitting model for each star in our sample in Figure~\ref{fig:SED} .

We find that the UV flux of WOCS 2068 can be well described by a combination of two stars -- one beginning to turn off the main-sequence, and the other a blue straggler of $\sim6800$ K. This SED fit is consistent with our interpretation of the photometric variability, that the system consists of one rapidly rotating merger or collision product (the 6800 K blue straggler), and the other a typical main-sequence star.  

Several other binaries in our sample also have UV excesses over single-star models that are not resolved by adding a main-sequence companion. These include WOCS 14020, 3001, 12020, 6025, and 11006.  

WOCS 14020 is the only system where we consider this UV excess a definitive white dwarf detection. Assuming a typical 0.5 \Msolar~C/O white dwarf model, the UV excess is most compatible with a $\sim13,000$ K white dwarf companion corresponding to an age of $\sim300$ Myr \citep{Tremblay2011}. 

The rest of the binaries with UV excesses are all hotter than WOCS 14020, making it more likely that the wide GALEX FUV passband picks up some flux from the Wien tail of the primary. Due to the low resolution and uncertainties on the spectral models in the UV, it is not clear whether these excesses indicate white dwarf companions. They could, for example, result from elevated UV flux due to chromospheric emission, which might be expected given that these stars are all rapidly rotating. Similar excesses have been discovered in other FGK field stars using GALEX photometry that have been largely attributed to UV emission from stellar activity \citep{Smith2014}. The excesses are nevertheless large enough to be intriguing. While none of the fluxes are large enough to indicate a very young white dwarf ($< 150$ Myr), they could be compatible with cooler white dwarf companions with ages of $\gtrsim150$ Myr. We suggest follow-up observations using more precise, multi-band UV photometry (e.g. HST/WFC3 as in \citealt{Gosnell2015}) to more definitively address the presence of white-dwarf companions.

Given the temperatures of the primaries and the GALEX detection limits, we might expect to detect white dwarf companions hotter than $\sim13000$ K  as UV excesses in the stellar SEDs, corresponding to an age younger than about $300$ Myr.  Notably, we detect UV excesses  only in  the binaries with the youngest rotational ages in our sample, all less than 300-400 Myr (Figure~\ref{fig:gyro}). The binary systems with older inferred gyro-ages (WOCS 4001, WOCS 9005) do not have UV detections. The gyro-ages and UV excesses are therefore both compatible with the hypothesis that 4001 and 9005 formed earlier, and therefore have cooler undetectable WD companions, and the other binaries formed recently enough to have hot, detectable WD companions. 

Only one of the single stars, WOCS 1020, has a GALEX FUV detection. The others, WOCS 2001 and WOCS 7035, do not have FUV detections. The non-detection of these stars is as expected, as only WOCS 1020 is hot enough to expect FUV flux above the GALEX detection limit. We show SEDs for all these systems in Figure~\ref{fig:SED}. All can be reasonably well fit with single-star SED models. 

\section{Summary and Discussion}
Based on the findings of \citet{Leiner2018}, we assume that mass transfer spins stars up to large rotational velocities at formation, and then these stars spin down as they age approximately as predicted by models for standard solar-type stars. As a result, a mass-transfer event restarts a star's gyro-age clock. Models and observations of other blue straggler formation mechanisms (mergers and collisions) also likely yield rapidly rotating stellar products, though their spindown behavior has not been empirically determined.  Thus it is not yet established that rotation rate is a reliable measure of time since formation for merger and collision products. 

This spin-up at formation enables identification of blue-straggler-type objects within the main sequence by looking for stars in old populations with fast rotation rates. We pilot the use of this technique in M67 by measuring stellar rotation periods from K2 Campaigns 5 and 16 light curves as well as from the spectral database of the WIYN Open Cluster Study. We find 11 rapid rotators on the M67 main sequence, all with rotation periods less than 8 days. None of these stars have close binary companions, so the rapid rotation is not due to tidal spin-up. We hypothesize that these stars have been through mass transfer, mergers, or stellar collisions within the past 1 Gyr.

These 11 detections have much in common with the blue straggler populations of old open clusters including: a binary fraction three times higher than the main-sequence spectroscopic binary fraction of M67; a preponderance of long-period, low-eccentricity binary systems; and in at least one case a UV excess indicative of a young white dwarf companion. These results suggest that these 11 sources are mass-transfer, merger, or collision products, and thus represent a low-luminosity extension of the blue straggler distribution. Given this hypothesis, we suggest calling this population of main-sequence blue stragglers ``blue lurkers," as they have been rejuvenated by mass-transfer/mergers like the blue stragglers, but blend with the normal main-sequence population in the color-magnitude diagram. Presumably as the cluster ages, similar-mass stars will begin to evolve away from the main-sequence, revealing these lurkers as the blue stragglers they truly are. 

Because these stars cannot be detected from photometry, this hidden population has not been characterized until now. With 11 detections, these blue lurkers are nearly as numerous as the classical blue straggler population of the cluster. This result suggests that population studies that focus only on classical blue stragglers are missing as much as half of the mass-transfer, merger and collision population. 

These 11 blue lurker detections were selected from a sample of $\sim400$ solar-type main-sequence cluster members. Thus at least 3\% of normal main-sequence stars are actually blue lurkers. In more detail, this sample includes 98 spectroscopic binaries (P$_\mathrm{orb} <$ 10$^4$ days; \citealt{Geller2015}, Geller et al. 2019, in prep). Eight of our 11 detections are among these binaries, implying at least 8\% of the solar-type spectroscopic binary population consists of blue lurkers. Of these 8, we suggest that at least 5 are recent mass-transfer products given their secondary masses (Figure~\ref{fig:wdperiod}), indicating at least 5\% of the solar-type spectroscopic binary population has been through mass transfer within the last Gyr. Three systems are observed to be single stars, indicating $\sim1$\% of the main sequence is composed of spun-up merger or collision products. If merger products spin down in much the same way as we have demonstrated for mass-transfer products, these stars may also have formed within the last Gyr. 

This result is illustrative of the ways binary evolution can impact a stellar population. If 11 stellar systems in M67 have been through a stellar collision or mass transfer event in the last Gyr, it is certain that other stars on the main sequence have also experienced stellar interactions in the more distant past and since spun down. Indeed, 4 of the 11 blue lurkers in our sample are near the zero-age main-sequence (Figure~\ref{CMD}). Such stars likely result from mass transfer onto lower mass, largely unevolved stars and themselves will live for several Gyr before evolving off the main-sequence. 

Older blue lurker systems would likely show up with intermediate rotation periods (i.e. 8-15 days). These slower rotators would be less magnetically active with smaller and shorter-lived spots, making them harder to identify through rotational modulation. Similarly, their slower rotation rates would make them undetectable using $v$sin$i$ measurements in all but very-high-resolution spectroscopic studies. These slower rotation periods are also closer to typical cluster rotation rates, making these blue lurkers more difficult to distinguish from typical M67 single stars. Such stars may nevertheless contribute to the spread of rotation rates observed in the cluster, skewing measurements towards shorter rotation periods and complicating efforts to calibrate precise gyrochronology relationships in clusters. 

As a simple upper bound, if we suppose formation rates of FGK-type blue lurkers remained constant over the 4 Gyr history of the cluster, this implies up to $\sim30$\% of solar-type binaries and $\sim4$\% of the single main-sequence stars have been through an interaction during their lifetimes. Though a very simple estimate, these numbers are in rough agreement with other studies. \citet{Andronov2006} predict 3-4\% of main-sequence stars in M67 may be merger products. \citet{Murphy2018} find $\sim$20\% of the field A/F-type binaries within a similar period range are post-mass-transfer binaries. The post-interaction fraction among older binary populations is sizable.

Several other clusters have the K2 and Kepler light curves needed to detect blue lurker populations including NGC 6791, NGC 6819, and Ruprecht 147. Studies of additional clusters such as these, combined with detailed binary-population modeling, will be required to quantify more precisely what the impact of binary interactions may be on the rotational properties of cluster stellar populations. 

The full population of blue lurkers on cluster main sequences remains largely unexplored because of the difficulties in detecting them. As a result, the stellar and orbital properties of this significant population have not been well characterized. Already there are hints that this population may yield new insights into binary evolution physics. For example, three stars in our sample have orbital periods of just 100-400 days, shorter periods than almost all blue straggler stars and an orbital period domain thought to be sparsely populated with main sequence-white dwarf binaries (e.g. \citealt{Willems2004}). For the two faintest systems, the inferred mass ratios ($\frac{M_\mathrm{donor}}{M_\mathrm{accretor}}$) at the onset of mass transfer strongly predict unstable mass transfer (e.g. \citealt{Chen2008}), yet they have survived without the dramatic orbital shrinkage expected during common envelope evolution. Building a larger sample of post-mass-transfer blue lurkers across more clusters is necessary to see if these types of orbits are indeed common among the class. Future detailed evolutionary modeling will also be required to better understand possible formation pathways.

Kepler and K2 have opened the door to detecting lower-luminosity mass-transfer, merger, and collision products on the main sequence using rotation rates. With similar future missions like TESS and PLATO planned for the near and longer term future, rotational studies of stars in clusters and in the field will continue to be important areas of study. TESS will yield rotation periods for nearby stars in younger clusters ($<1$ Gyr) and the field. Looking for rapidly rotating field stars with abundances or kinematics indicative of old age may be a viable detection method for field post-mass-transfer binaries.  Due to the large scatter in rotation rates among young stars, this technique may not be well suited to identifying interaction products in TESS clusters.  If PLATO, planned for launch in 2026, targets more older clusters, more rotational identifications of blue lurkers may be possible. As the known population continues to grow, these populations can provide new tests for binary evolution physics and cluster population models.

\acknowledgements{EL is supported by an NSF Astronomy and Astrophysics Postdoctoral Fellowship under award AST-1801937. AV's work was performed under contract with the California Institute of Technology (Caltech)/Jet Propulsion Laboratory (JPL) funded by NASA through the Sagan Fellowship Program executed by the NASA Exoplanet Science Institute.

\end{document}